# Reversible "triple-Q" elastic field structures in a chiral magnet

Yangfan Hu[1,*], Biao Wang[1,**]

[1]Sino-French Institute of Nuclear Engineering and Technology, Sun Yat-Sen University, 510275 GZ, China

The analytical solution of the periodic elastic fields in chiral magnets caused by presence of periodically distributed eigenstrains is obtained. For the skyrmion phase, both the periodic displacement field and the stress field are composed of three "triple-Q" structures with different wave numbers. The periodic displacement field, obtained by combining the three "triple-Q" displacement structures, is found to have the same lattice vectors with the magnetic skyrmion lattice. We find that for increasing external magnetic field, one type of "triple-Q" displacement structure and stress structure undergo a "configurational reversal", where the initial and the final field configuration share similar pattern but with opposite direction of all the field vectors. The solution obtained is of fundamental significance for understanding the emergent mechanical properties of skyrmions in chiral magnets.

**Introduction**

Chiral magnets have attracted interest over the last few years due to experimental observation of a new chiral modulated magnetic state, commonly referred to as skyrmion lattice, first in $MnSi$[1], and then in $Fe_{0.5}Co_{0.5}Si$[2] and $FeGe$[3]. Skyrmion lattice in chiral magnets can be described as chiral spin structures with a whirling configuration, which can be described mathematically by a combination of three plane-wave functions in space (hence referred to as a "triple-Q" structure). These magnetic skyrmions are stabilized by the antisymmetric Dzyaloshinskii-Moriya (DM) interactions[4,5], and are well known for their emergent electromagnetic properties[6,7] and topological Hall effect [8,9].

Due to the magnetoelastic coupling in chiral magnets, it is known that application of mechanical loads or misfit strains can stabilize the skyrmion lattice[2,10-12]. Moreover, it is found in $FeGe$ that elastic deformation and the deformation of the skyrmion lattice are strongly coupled, which leads to large emergent deformation of skyrmion lattice when elastic stress is applied[13,14]. To understand such an exotic phenomenon, we first have to discuss the internal elastic field induced by presence of magnetic skyrmions. This induced elastic field should be composed of a homogeneous part, which has been solved in our previous work[15], and a periodic part, which is to be discussed in the present work. In the early studies of MnSi, it has already been confirmed that periodic distribution of elastic fields coexists with periodic magnetization in the spin-density-wave phases (e.g., helical and conical phase)[16]. Besides, it is shown theoretically that the magnitude of periodic strain waves should be considered as independent variables when formulating the free energy of the system, since it results in new terms in the free energy functional[17]. Existence of the periodic elastic field represents realization of the same mathematical structure as magnetic skyrmions in a different physical field due to multiphysics coupling. Moreover, regarding the nonlinear nature of the magnetoelastic coupling in chiral magnets, additional periodic structures with changed wave vectors may occur in the solution of elastic field. Induced by the magnetic skyrmions, these periodic elastic structures should always accompany the magnetic structures. Hence the solution of the periodic

*Corresponding author. E-mail: huyf3@mail.sysu.edu.cn (Yangfan Hu),
**Corresponding author. E-mail: wangbiao@mail.sysu.edu.cn (Biao Wang).

elastic fields may characterize the elastic property of magnetic skyrmions in some way.

The elasticity problem induced by presence of any kind of phase transition strains is called an eigenstrain problem [16] in micromechanics. In this case, the eigenstrains refer to the strains that occur due to a change of magnetization during a phase transition from the paramagnetic phase to the skyrmion phase. Since in the skyrmion phase, the magnetization is a periodic vector function in space, we encounter an eigenstrain problem with periodic eigenstrains. The analytical method for solving the elastic fields induced by periodic distribution of eigenstrains was developed long ago [18,19], mainly due to its mathematical significance for constructing the Fourier transform-based analytical solution method. Yet, the eigenstrain problem with periodic eigenstrains composed of plane waves with several discrete wave vectors was not treated. Before the discovery of magnetic skyrmions in chiral magnets, it's not clear such a solution is of any physical significance.

In this paper, we formulate the eigenstrain problem from an extended micromagnetic model for chiral magnets developed upon group theoretical analysis and the Ginzburg-Landau theory[15], and then obtain the analytical solution of the problem for two different chiral magnetic phases: the conical phase and the skyrmion phase. We find that appearance of the triple-Q skyrmion lattices is accompanied by formation of three types of triple-Q structures in the displacement field, described by $\mathbf{u}^{S1}(\mathbf{r})$ (with the same wave number $q$ as the magnetic skyrmions), $\mathbf{u}^{S2}(\mathbf{r})$ (wave number $2q$), and $\mathbf{u}^{S3}(\mathbf{r})$ (wave number $\sqrt{3}q$), as well as formation of three types of triple-Q structures in the stress field, described by $\sigma_{ij}^{S1}$ (wave number $q$), $\sigma_{ij}^{S2}$ (wave number $2q$), and $\sigma_{ij}^{S3}$ (wave number $\sqrt{3}q$). By using the values of equilibrium magnetization obtained through free energy minimization at given temperature and magnetic field[15], we plot the field configuration of $\mathbf{u}^{S1}(\mathbf{r})$, $\mathbf{u}^{S2}(\mathbf{r})$, $\mathbf{u}^{S3}(\mathbf{r})$ and $\sigma_{3j}^{S1}$, $\sigma_{3j}^{S2}$, $\sigma_{3j}^{S3}$ and discuss their variation with temperature and magnetic field. We find that as the applied magnetic field increases, the field configuration of $\mathbf{u}^{S1}(\mathbf{r})$ and $\sigma_{3j}^{S1}$ gradually undergoes a "configurational reversal" process, where the field configuration of the initial state and the final state remains similar, but the direction of every vector in the field is reversed. The phenomenon provides a possibility of developing novel information storage devises and microwave applications. The analytical solution of the periodic elastic fields also lay a foundation for discussion of the emergent elastic behavior of magnetic skyrmions[13], which is to be discussed in a subsequent work of ours.

**Formulation of the eigenstrain problem in chiral magnets**

For chiral magnetics with cubic symmetry, the Helmholtz free energy density contains two terms related to the elastic strains[15], which are the elastic energy density

$$w_{el} = \frac{1}{2}C_{11}(\varepsilon_{11}^2 + \varepsilon_{22}^2 + \varepsilon_{33}^2) + C_{12}(\varepsilon_{11}\varepsilon_{22} + \varepsilon_{11}\varepsilon_{33} + \varepsilon_{22}\varepsilon_{33}) \\ + \frac{1}{2}C_{44}(\gamma_{12}^2 + \gamma_{13}^2 + \gamma_{23}^2), \quad (1)$$

and the magnetoelastic energy density

$$w_{me} = \frac{1}{M_s^2}[L_1(M_1^2\varepsilon_{11} + M_2^2\varepsilon_{22} + M_3^2\varepsilon_{33}) + L_2(M_3^2\varepsilon_{11} + M_1^2\varepsilon_{22} + M_2^2\varepsilon_{33}) \\ + L_3(M_1M_2\gamma_{12} + M_1M_3\gamma_{13} + M_2M_3\gamma_{23}) + KM^2\varepsilon_{ii} + \sum_{i=1}^{6} L_{0i}f_{0i}], \quad (2)$$

where $C_{11}$, $C_{12}$, and $C_{44}$ are the elastic stiffness at constant magnetization, $\varepsilon_{ij}$ ($i,j = 1,2,3$) are the elastic strains, $\gamma_{12} = 2\varepsilon_{12}, \gamma_{13} = 2\varepsilon_{13}$, and $\gamma_{23} = 2\varepsilon_{23}$ are the engineering shear strains, $M_s$ denotes the saturation magnetization, $M^2 = M_1^2 + M_2^2 + M_3^2$ and

$$\begin{aligned}
f_{O1} &= \varepsilon_{11}(M_{1,2}M_3 - M_{1,3}M_2) + \varepsilon_{22}(M_{2,3}M_1 - M_{2,1}M_3) + \varepsilon_{33}(M_{3,1}M_2 - M_{3,2}M_1), \\
f_{O2} &= \varepsilon_{11}(M_{3,1}M_2 - M_{2,1}M_3) + \varepsilon_{22}(M_{1,2}M_3 - M_{3,2}M_1) + \varepsilon_{33}(M_{2,3}M_1 - M_{1,3}M_2), \\
f_{O3} &= \varepsilon_{11}M_1(M_{2,3} - M_{3,2}) + \varepsilon_{22}M_2(M_{3,1} - M_{1,3}) + \varepsilon_{33}M_3(M_{1,2} - M_{2,1}), \\
f_{O4} &= \gamma_{23}(M_{1,3}M_3 - M_{1,2}M_2) + \gamma_{13}(M_{2,1}M_1 - M_{2,3}M_3) + \gamma_{12}(M_{3,2}M_2 - M_{3,1}M_1), \\
f_{O5} &= \gamma_{23}(M_{3,1}M_3 - M_{2,1}M_2) + \gamma_{13}(M_{1,2}M_1 - M_{3,2}M_3) + \gamma_{12}(M_{2,3}M_2 - M_{1,3}M_1), \\
f_{O6} &= \gamma_{23}M_1(M_{3,3} - M_{2,2}) + \gamma_{13}M_2(M_{1,1} - M_{3,3}) + \gamma_{12}M_3(M_{2,2} - M_{1,1}).
\end{aligned} \quad (3)$$

In eq. (2), higher order terms of magnetoelastic coupling (e.g., $w_{me2}$ in Ref.[15]) are omitted for convenience of deduction. According to our previous work[15], such a simplification leads to an estimated error in the order of 0.01% when calculating the elastic strains $\varepsilon_{23}, \varepsilon_{13}$ and $\varepsilon_{12}$ in $MnSi$.

For a bulk material free from any mechanical loads, the stresses are obtained from $\sigma_{ij} = \frac{\partial(w_{el}+w_{me})}{\partial \varepsilon_{ij}}$ when $i = j$, and $\sigma_{ij} = \frac{\partial(w_{el}+w_{me})}{\partial \gamma_{ij}}$ when $i \neq j$. After manipulation we get

$$\begin{aligned}
\sigma_{11} &= C_{11}(\varepsilon_{11} - \varepsilon_{11}^*) + C_{12}(\varepsilon_{22} - \varepsilon_{22}^* + \varepsilon_{33} - \varepsilon_{33}^*), \\
\sigma_{22} &= C_{11}(\varepsilon_{22} - \varepsilon_{22}^*) + C_{12}(\varepsilon_{11} - \varepsilon_{11}^* + \varepsilon_{33} - \varepsilon_{33}^*), \\
\sigma_{33} &= C_{11}(\varepsilon_{33} - \varepsilon_{33}^*) + C_{12}(\varepsilon_{11} - \varepsilon_{11}^* + \varepsilon_{22} - \varepsilon_{22}^*), \\
\sigma_{23} &= C_{44}(\gamma_{23} - \gamma_{23}^*), \\
\sigma_{13} &= C_{44}(\gamma_{13} - \gamma_{13}^*), \\
\sigma_{12} &= C_{44}(\gamma_{12} - \gamma_{12}^*),
\end{aligned} \quad (4)$$

where $\varepsilon_{ij}^*$ and $\gamma_{ij}^*$, the eigenstrains, are related to the magnetization:

$$\begin{aligned}
\varepsilon_{11}^* &= K^*M^2 - L_1^*M_1^2 - L_2^*M_3^2 + L_{O1}^*(M_3M_{1,2} - M_2M_{1,3}) + L_{O2}^*(M_3M_{2,1} - M_2M_{3,1}) + L_{O3}^*M_1(M_{2,3} - M_{3,2}), \\
\varepsilon_{22}^* &= K^*M^2 - L_1^*M_2^2 - L_2^*M_1^2 + L_{O1}^*(M_1M_{2,3} - M_3M_{2,1}) + L_{O2}^*(M_1M_{3,2} - M_3M_{1,2}) + L_{O3}^*M_2(M_{3,1} - M_{1,3}), \\
\varepsilon_{33}^* &= K^*M^2 - L_1^*M_3^2 - L_2^*M_2^2 + L_{O1}^*(M_2M_{3,1} - M_1M_{3,2}) + L_{O2}^*(M_2M_{1,3} - M_1M_{2,3}) + L_{O3}^*M_3(M_{1,2} - M_{2,1}), \\
\gamma_{23}^* &= \frac{-L_3M_2M_3 + L_{O6}M_1(M_{2,2} - M_{3,3}) + M_2(L_{O4}M_{1,2} + L_{O5}M_{2,1}) - M_3(L_{O4}M_{1,3} + L_{O5}M_{3,1})}{C_{44}M_s^2}, \\
\gamma_{13}^* &= \frac{-L_3M_1M_3 + L_{O6}M_2(M_{3,3} - M_{1,1}) + M_3(L_{O4}M_{2,3} + L_{O5}M_{3,2}) - M_1(L_{O4}M_{2,1} + L_{O5}M_{1,2})}{C_{44}M_s^2}, \\
\gamma_{12}^* &= \frac{-L_3M_1M_2 + L_{O6}M_3(M_{1,1} - M_{2,2}) + M_1(L_{O4}M_{3,1} + L_{O5}M_{1,3}) - M_2(L_{O4}M_{3,2} + L_{O5}M_{2,3})}{C_{44}M_s^2}.
\end{aligned} \quad (5)$$

In eq. (5), the parameters with a superscript "*" are defined as $K^* = \frac{-C_{11}K + C_{12}(K+L_1+L_2)}{(C_{11}-C_{12})(C_{11}+2C_{12})M_s^2}$, $L_1^* = \frac{L_1}{(C_{11}-C_{12})M_s^2}$, $L_2^* = \frac{L_2}{(C_{11}-C_{12})M_s^2}$, $L_{O1}^* = \frac{-C_{11}L_{O1} + C_{12}(-L_{O1}+L_{O2}+L_{O3})}{(C_{11}-C_{12})(C_{11}+2C_{12})M_s^2}$, $L_{O2}^* = \frac{C_{11}L_{O2} - C_{12}(L_{O1}-L_{O2}+L_{O3})}{(C_{11}-C_{12})(C_{11}+2C_{12})M_s^2}$, $L_{O3}^* = \frac{C_{12}(L_{O1}+L_{O2}) - (C_{11}+C_{12})L_{O3}}{(C_{11}-C_{12})(C_{11}+2C_{12})M_s^2}$.

Substituting eq. (4) into the equilibrium equation, and using the geometrical equations under small-deformation assumption, we have

$$\begin{aligned}
C_{11}u_{1,11} + C_{44}(u_{1,22} + u_{1,33}) + (C_{12} + C_{44})(u_{2,12} + u_{3,13}) &= X_1, \\
C_{11}u_{2,22} + C_{44}(u_{2,11} + u_{2,33}) + (C_{12} + C_{44})(u_{1,12} + u_{3,23}) &= X_2, \\
C_{11}u_{3,33} + C_{44}(u_{3,22} + u_{3,11}) + (C_{12} + C_{44})(u_{2,23} + u_{1,13}) &= X_3,
\end{aligned} \quad (6)$$

where

$$X_1 = C_{11}\varepsilon^*_{11,1} + C_{12}(\varepsilon^*_{22,1} + \varepsilon^*_{33,1}) + C_{44}(\gamma^*_{12,2} + \gamma^*_{13,3}),$$
$$X_2 = C_{11}\varepsilon^*_{22,2} + C_{12}(\varepsilon^*_{11,2} + \varepsilon^*_{33,2}) + C_{44}(\gamma^*_{12,1} + \gamma^*_{23,3}), \quad (7)$$
$$X_3 = C_{11}\varepsilon^*_{33,3} + C_{12}(\varepsilon^*_{22,3} + \varepsilon^*_{11,3}) + C_{44}(\gamma^*_{23,2} + \gamma^*_{13,1}).$$

Here $X_i$ ($i = 1, 2, 3$) resemble components of the body force caused by presence of eigenstrains.

Assume $\mathbf{M}(\mathbf{r})$ and $\mathbf{u}(\mathbf{r})$ to be periodic functions in space, which can be expressed as

$$\mathbf{M}(\mathbf{r}) = \sum_{\mathbf{q}} \mathbf{m}^{\mathbf{q}} e^{i\mathbf{q}\cdot\mathbf{r}},$$
$$\mathbf{u}(\mathbf{r}) = \sum_{\mathbf{q}} \mathbf{U}^{\mathbf{q}} e^{i\mathbf{q}\cdot\mathbf{r}}, \quad (8)$$

where $\mathbf{m}^{-\mathbf{q}} = \hat{\mathbf{m}}^{\mathbf{q}}$ and $\mathbf{U}^{-\mathbf{q}} = \hat{\mathbf{U}}^{\mathbf{q}}$. Here $\hat{\mathbf{m}}^{\mathbf{q}}$ and $\hat{\mathbf{U}}^{\mathbf{q}}$ are complex conjugates of $\mathbf{m}^{\mathbf{q}}$ and $\mathbf{U}^{\mathbf{q}}$. Through eq. (8), the equilibrium state of the system is determined by minimizing $F$ with respect to $\mathbf{m}^{\mathbf{q}}$, $\mathbf{U}^{\mathbf{q}}$ and $\mathbf{q}$. When the system is free from external mechanical loads, the solution of $\mathbf{U}^{\mathbf{q}}$ can be obtained by solving the eigenstrain problem for chiral magnets as

$$\mathbf{U}^{\mathbf{q}} = \mathbf{K}_{\mathbf{q}}^{-1} \mathbf{X}^{\mathbf{q}}, \quad (9)$$

where

$$\mathbf{K}_{\mathbf{q}} = -\begin{bmatrix} C_{11}q_1^2 + C_{44}(q_2^2 + q_3^2) & (C_{11} + C_{44})q_1 q_2 & (C_{11} + C_{44})q_1 q_3 \\ (C_{11} + C_{44})q_1 q_2 & C_{11}q_2^2 + C_{44}(q_1^2 + q_3^2) & (C_{11} + C_{44})q_2 q_3 \\ (C_{11} + C_{44})q_1 q_3 & (C_{11} + C_{44})q_2 q_3 & C_{11}q_3^2 + C_{44}(q_1^2 + q_2^2) \end{bmatrix}, \quad (10)$$

and $\mathbf{X}^{\mathbf{q}}$ is solved from $\mathbf{X} = \sum_{\mathbf{q}} \mathbf{X}^{\mathbf{q}} e^{i\mathbf{q}\cdot\mathbf{r}}$. Eq. (9) determines the elastic fields in the material when the magnetic state of the system is determined.

Before moving on to the solution of eq. (6) for different chiral magnetic phases, we give a brief discussion on the situation in the ferromagnetic phase. When the material is stabilized in a ferromagnetic state, the magnetization $\mathbf{M}$ is a constant vector inside the material. In this case, the eigenstrains are constants according to their definition given in eq. (5), and we have $X_i = 0$ ($i = 1, 2, 3$) from eq. (7). We thus obtain a solution of zero stresses for free boundary condition. The physical interpretation of this solution is clear: when the material is free from external loads, it is free to deform and so the total strains equal to the eigenstrains, while the stresses vanish.

**Solution for the conical phase**

In the conical phase, the magnetization can be written without loss of generality as

$$\mathbf{M} = [m_q \cos(qz), m_q \sin(qz), m_3]^T, \quad (11)$$

where it is assumed that the magnetic field is applied along the z-axis. Of course, the following results change with the direction of the magnetic field. But the method and the form of the solution are exactly the same. Substitution of eq. (11) into eq. (5) yields

$$\varepsilon^*_{ij} = \varepsilon^{c*}_{ij} + \mathrm{Re}[\varepsilon^{q1*}_{ij} e^{iqz} + \varepsilon^{q2*}_{ij} e^{2iqz}], \quad (12)$$

where $\varepsilon^*_{ij} = \frac{1}{2}\gamma^*_{ij}, i \neq j$, and

$$\varepsilon_{11}^{c*} = \left[K^* + \frac{1}{2}(-L_1^* + L_{O1}^*q + L_{O3}^*q)m_q^2 - L_2^*m_3^2\right],$$

$$\varepsilon_{22}^{c*} = \left[K^* + \frac{1}{2}(-L_1^* - L_2^* + L_{O1}^*q + L_{O3}^*q)m_q^2\right], \quad (13)$$

$$\varepsilon_{33}^{c*} = \left(K^* - L_1^*m_3^2 - L_{O2}^*qm_q^2 - \frac{1}{2}L_2^*m_q^2\right),$$

$$\varepsilon_{ij}^{c*} = 0 \quad \text{when } i \neq j,$$

$$[\varepsilon_{ij}^{q1*}] = -\frac{m_q m_3 (L_3 - L_{O4}q)}{2C_{44}M_s^2}\begin{bmatrix} 0 & 0 & 1 \\ 0 & 0 & -i \\ 1 & -i & 0 \end{bmatrix}, \quad (14)$$

$$[\varepsilon_{ij}^{q2*}] = \frac{m_q^2}{2}\begin{bmatrix} -(L_1^* + L_{O1}^*q - L_{O3}^*q) & \dfrac{i(L_3 + 2L_{O5}q)}{2C_{44}M_s^2} & 0 \\ \dfrac{i(L_3 + 2L_{O5}q)}{2C_{44}M_s^2} & L_1^* - L_2^* + L_{O1}^*q - L_{O3}^*q & 0 \\ 0 & 0 & L_2^* \end{bmatrix}, \quad (15)$$

and $i = \sqrt{-1}$. The solution of the displacement field can be sought in the following form

$$\mathbf{u} = \mathbf{u}^c + \text{Re}[\mathbf{u}^{q1}e^{iqz} + \mathbf{u}^{q2}e^{2iqz}], \quad (16)$$

where $\mathbf{u}^c$ dentoes the displacement field which corresponds to constant eigenstrains. By direct integration of eq. (13), we have

$$\begin{aligned} u_1^c &= \varepsilon_{11}^{c*}x, \\ u_2^c &= \varepsilon_{22}^{c*}y, \\ u_3^c &= \varepsilon_{33}^{c*}z, \end{aligned} \quad (17)$$

where rigid body movements are not considered. Substituting eqs. (14-16) into eqs. (6, 7), after manipulation we obtain

$$\begin{aligned} \mathbf{u}^{q1} &= \mathbf{K}^{-1}\mathbf{X}^{q1}, \\ \mathbf{u}^{q2} &= \frac{1}{4}\mathbf{K}^{-1}\mathbf{X}^{q2}, \end{aligned} \quad (18)$$

where

$$\mathbf{K} = \begin{bmatrix} C_{44}q^2 & 0 & 0 \\ 0 & C_{44}q^2 & 0 \\ 0 & 0 & C_{11}q^2 \end{bmatrix},$$

$$\mathbf{X}^{q1} = \frac{m_q m_3}{M_s^2}(L_3 - L_{O4}q)q[i, 1, 0]^T, \quad (19)$$

$$\mathbf{X}^{q2} = \left[0, 0, -i(C_{11} - C_{12})L_2^*qm_q^2\right]^T.$$

Eq. (18) can thus be expanded as

$$\begin{aligned} \mathbf{u}^{q1} &= \frac{m_q m_3 (L_3 - L_{O4}q)}{C_{44}qM_s^2}[i, 1, 0]^T, \\ \mathbf{u}^{q2} &= \left[0, 0, \frac{-i(C_{11} - C_{12})L_2^*m_q^2}{4C_{11}q}\right]^T. \end{aligned} \quad (20)$$

The solution for the displacement field can be obtained by combining eqs. (16, 17, 20). The total strains and stresses are solved as

$$[\varepsilon_{ij}] = [\varepsilon_{ij}^{c*}] - \frac{m_q m_3 (L_3 - L_{O4}q)}{2C_{44}M_s^2} \begin{bmatrix} 0 & 0 & \cos(qz) \\ 0 & 0 & \sin(qz) \\ \cos(qz) & \sin(qz) & 0 \end{bmatrix}$$
$$+ \frac{(C_{11} - C_{12})L_2^* m_q^2}{2C_{11}} \begin{bmatrix} 0 & 0 & 0 \\ 0 & 0 & 0 \\ 0 & 0 & \cos(2qz) \end{bmatrix}, \quad (21)$$

and

$$[\sigma_{ij}] = \frac{m_q^2}{2M_s^2} \begin{bmatrix} [L_1 + \frac{C_{12}}{C_{11}}L_2 - (L_{O1} - L_{O3})q]\cos(2qz) & (L_3 + 2L_{O5}q)\sin(2qz) & 0 \\ (L_3 + 2L_{O5}q)\sin(2qz) & -[L_1 - \frac{C_{11}+C_{12}}{C_{11}}L_2 - (L_{O1} - L_{O3})q]\cos(2qz) & 0 \\ 0 & 0 & 0 \end{bmatrix}. \quad (22)$$

In eq. (12) the periodic eigenstrains contain a part with a wave number $2q$. This wave-number-doubling phenomenon derives from the nonlinear nature of the magnetostriction effect, and has been discussed before[16]. By solving the elastic fields of this eigenstrain problem, it is shown that the magnitude of the periodic part of stresses with wave number $q$ vanishes. To explains the physical origin of this result, one needs to examine the compatibility condition for the eigenstrains $\varepsilon_{ij}^*$. For the three composition of $\varepsilon_{ij}^*$ in eq. (12), $\varepsilon_{ij}^{c*}$ and $\mathbf{Re}[\varepsilon_{ij}^{q1*}e^{iqz}]$ satisfy the compatibility condition, while a part of $\mathbf{Re}[\varepsilon_{ij}^{q2*}e^{2iqz}]$ does not satisfy the compatibility condition. The compatible part of eigenstrains directly generates elastic strains, while the incompatible part of eigenstrains is constrained by the elastic body through an internal stress field given in eq. (22).

Using the related parameters for $MnSi$:

$$C_{11} = 283.3 \text{ GPa}, C_{12} = 64.1 \text{ GPa}, C_{44} = 117.9 \text{ GPa }^7,$$

$$K = -2 \times 10^7 \text{ JA}^{-2}\text{m}^{-1}, L_1 = -0.7 \times 10^6 \text{ JA}^{-2}\text{m}^{-1}, L_2 = 0.6 \times 10^6 \text{ JA}^{-2}\text{m}^{-1}, L_3 = 1.646 \times 10^6 \text{ JA}^{-2}\text{m}^{-1}, ^{15}$$

$$L_{O1} = 1.147 \times 10^{-4} \text{ JA}^{-2}\text{m}^{-2}, L_{O2} = -0.573 \times 10^{-4} \text{ JA}^{-2}\text{m}^{-2}, L_{O3} = -0.573 \times 10^{-4} \text{ JA}^{-2}\text{m}^{-2}, ^{15}$$

$$L_{O4} = L_{O5} = L_{O6} = 0, ^{15}$$

it is found that for $|\mathbf{M}| = M_s$, $|\mathbf{u}^{p1}|_{m_3=m_q} \approx 3.1 \times 10^{-5}$nm, $|\mathbf{u}^{p2}|_{m_3=m_q} \approx 5.88 \times 10^{-7}$nm.

**Solution for the skyrmion phase**

In the skyrmion phase, the magnetization vector can be written as

$$\mathbf{M} = \begin{bmatrix} 0 \\ 0 \\ m_3 \end{bmatrix} + \frac{\sqrt{3}m_q}{3} \left\{ \begin{bmatrix} 0 \\ \sin(\mathbf{q}_1\mathbf{r}) \\ -\cos(\mathbf{q}_1\mathbf{r}) \end{bmatrix} + \begin{bmatrix} -\frac{\sqrt{3}}{2}\sin(\mathbf{q}_2\mathbf{r}) \\ -\frac{1}{2}\sin(\mathbf{q}_2\mathbf{r}) \\ -\cos(\mathbf{q}_2\mathbf{r}) \end{bmatrix} + \begin{bmatrix} \frac{\sqrt{3}}{2}\sin(\mathbf{q}_3\mathbf{r}) \\ -\frac{1}{2}\sin(\mathbf{q}_3\mathbf{r}) \\ -\cos(\mathbf{q}_3\mathbf{r}) \end{bmatrix} \right\}, \quad (23)$$

where it is assumed that the external magnetic field is applied along the z-axis, and $\mathbf{q}_1 = q[1,0,0]^T$, $\mathbf{q}_2 = q[-\frac{1}{2}, \frac{\sqrt{3}}{2}, 0]^T$, $\mathbf{q}_3 = q[-\frac{1}{2}, -\frac{\sqrt{3}}{2}, 0]^T$, $\mathbf{r} = [x, y, z]^T$. For the magnetization defined in eq. (23), the solution of the displacement field can be sought in the following form

$$\mathbf{u} = \mathbf{u}^c + \mathbf{Re}[\mathbf{u}^{a1}e^{i\mathbf{q}_1\mathbf{r}} + \mathbf{u}^{b1}e^{i\mathbf{q}_2\mathbf{r}} + \mathbf{u}^{c1}e^{i\mathbf{q}_3\mathbf{r}} + \mathbf{u}^{a2}e^{2i\mathbf{q}_1\mathbf{r}} + \mathbf{u}^{b2}e^{2i\mathbf{q}_2\mathbf{r}} + \mathbf{u}^{c2}e^{2i\mathbf{q}_3\mathbf{r}}$$
$$+ \mathbf{u}^{ab1}e^{i(\mathbf{q}_1-\mathbf{q}_2)\mathbf{r}} + \mathbf{u}^{ac1}e^{i(\mathbf{q}_1-\mathbf{q}_3)\mathbf{r}} + \mathbf{u}^{bc1}e^{i(\mathbf{q}_2-\mathbf{q}_3)\mathbf{r}}]. \quad (24)$$

One should notice that $\mathbf{q}_1 + \mathbf{q}_2 + \mathbf{q}_3 = 0$, and thus terms with $e^{i(\mathbf{q}_1+\mathbf{q}_2)\mathbf{r}}$, $e^{i(\mathbf{q}_1+\mathbf{q}_3)\mathbf{r}}$ and $e^{i(\mathbf{q}_2+\mathbf{q}_3)\mathbf{r}}$ are merged with terms with $e^{i\mathbf{q}_3\mathbf{r}}$, $e^{i\mathbf{q}_2\mathbf{r}}$ and $e^{i\mathbf{q}_1\mathbf{r}}$, respectively. By substituting eqs. (23, 24) into eqs. (5, 6, 7), the solution for the displacement field can be obtained in the same way as above. We have after manipulation

$$u_1^C = \left[K^*(m_q^2 + m_3^2) + \frac{1}{4}(-L_1^* - 2L_2^* + L_{O1}^*q - 2L_{O2}^*q + L_{O3}^*q)m_q^2 - L_2^* m_3^2\right]x,$$

$$u_2^C = \left[K^*(m_q^2 + m_3^2) + \frac{1}{4}(-L_1^* - L_2^* + L_{O1}^*q - 2L_{O2}^*q + L_{O3}^*q)m_q^2\right]y, \quad (25)$$

$$u_3^C = \left[K^*(m_q^2 + m_3^2) + \frac{1}{4}(-2L_1^* - L_2^* + 2L_{O1}^*q + 2L_{O3}^*q)m_q^2 - L_1^* m_3^2\right]z,$$

$$\mathbf{u}^{a1} = \frac{im_q}{12qC_{11}(C_{11} - C_{12})M_s^2}(g_c^{a1}m_3 + g_s^{a1}m_q)[1,0,0]^T, \quad (26)$$

$$u_1^{b1} = -\frac{im_q}{12q(C_{11} - C_{12})C_kM_s^2}(g_{c1}^{b1}m_3 + g_{s1}^{b1}m_q),$$

$$u_2^{b1} = \frac{im_q}{12q(C_{11} - C_{12})C_kM_s^2}(g_{c2}^{b1}m_3 + g_{s2}^{b1}m_q), \quad (27)$$

$$u_3^{b1} = -\frac{m_q^2(L_{O4} + L_{O5} + L_{O6})}{16\sqrt{3}C_{44}M_s^2},$$

$$\begin{aligned} u_1^{c1} &= u_1^{b1}, \\ u_2^{c1} &= -u_2^{b1}, \\ u_3^{c1} &= -u_3^{b1}, \end{aligned} \quad (28)$$

$$\mathbf{u}^{a2} = \frac{im_q^2 L_2}{12qC_{11}M_s^2}[1,0,0]^T, \quad (29)$$

$$u_1^{b2} = -\frac{im_q^2}{24qC_kM_s^2}[13C_{44}L_2 + 3C_{12}(L_1 + 3L_2 - L_3 - qL_{O1} + qL_{O3} + 2qL_{O4} + 2qL_{O6})$$
$$-3C_{11}(3L_1 - 4L_2 - 3L_3 - 3qL_{O1} + 3qL_{O3} + 6qL_{O4} + 6qL_{O6})],$$

$$u_2^{b2} = -\frac{im_q^2}{24qC_kM_s^2}[13C_{44}L_2 + C_{12}(-3L_1 + 4L_2 + 3L_3 + 3qL_{O1} - 3qL_{O3} - 6qL_{O4} - 6qL_{O6}) \quad (30)$$
$$+C_{11}(L_1 + 3L_2 - L_3 - qL_{O1} + qL_{O3} + 2qL_{O4} + 2qL_{O6})],$$

$$u_3^{b2} = -\frac{m_q^2(L_{O4} + L_{O5} + L_{O6})}{32\sqrt{3}C_{44}M_s^2},$$

$$\begin{aligned} u_1^{c2} &= u_1^{b2}, \\ u_2^{c2} &= -u_2^{b2}, \\ u_3^{c2} &= -u_3^{b2}, \end{aligned} \quad (31)$$

$$u_1^{ab1} = \frac{im_q^2}{12q(C_{11} - C_{12})C_kM_s^2}\{C_{11}^2(4K + 8L_2 + 2L_3 + 3qL_{O1} + qL_{O2} - 4qL_{O4} - 2qL_{O6})$$
$$+C_{12}[-4C_{44}(2K + L_1 + 4L_2 + qL_{O1} + qL_{O3}) + C_{12}(4K - 4L_1 - 4L_2 + 6L_3 + 5qL_{O1}$$
$$+3qL_{O2} - 4qL_{O3} - 12qL_{O4} - 6qL_{O6})] - 4C_{11}[-C_{44}(2K + L_1 + 6L_2 + qL_{O1} + qL_{O3})$$
$$+C_{12}(2K - L_1 + L_2 + 2L_3 + 2qL_{O1} + qL_{O2} - qL_{O3} - 4qL_{O4} - 2qL_{O6})]\},$$

$$u_2^{ab1} = -\frac{im_q^2}{12\sqrt{3}q(C_{11} - C_{12})C_kM_s^2}\{3C_{11}^2[4K - 4L_1 + 6L_3 + q(5L_{O1} + 3L_{O2} - 4L_{O3} - 12L_{O4} \quad (32)$$
$$-6L_{O6})] + C_{12}[4C_{44}(2K + L_1 + 4L_2 + qL_{O1} + qL_{O3}) + 3C_{12}(4K + 4L_2 + 2L_3$$
$$+3qL_{O1} + qL_{O2} - 4qL_{O4} - 2qL_{O6})] - 4C_{11}[C_{44}(2K + L_1 + 6L_2 + qL_{O1} + qL_{O3})$$
$$+3C_{12}(2K - L_1 + L_2 + 2L_3 + 2qL_{O1} + qL_{O2} - qL_{O3} - 4qL_{O4} - 2qL_{O6})]\},$$

$$u_3^{ab1} = -\frac{m_q^2(L_{O4} + L_{O5} + L_{O6})}{16\sqrt{3}C_{44}M_s^2},$$

$$u_1^{ac1} = u_1^{ab1},$$
$$u_2^{ac1} = -u_2^{ab1}, \tag{33}$$
$$u_3^{ac1} = -u_3^{ab1},$$

$$\mathbf{u}^{bc1} = \frac{im_q^2[(C_{11} - C_{12})(2K + L_1 - 5L_2 + qL_{O1} + qL_{O3}) + 2C_{11}L_2]}{12\sqrt{3}q(C_{11} - C_{12})C_{11}M_s^2}[0,1,0]^T, \tag{34}$$

where

$$g_c^{a1} = -4\sqrt{3}\big(-C_{12}(2K + qL_{O2}) + C_{11}(2K + 2L_2 + qL_{O2})\big),$$
$$g_s^{a1} = (-C_{12}(6K + 3L_1 - 2L_2 + 3qL_{O1} + 3qL_{O3}) \tag{35}$$
$$+C_{11}(6K + 3L_1 + 4L_2 + 3qL_{O1} + 3qL_{O3})),$$

$$g_{c1}^{b1} = -2\sqrt{3}\{-4C_{11}\{2C_{44}(2K - L_2 + qL_{O2}) + 3C_{12}[4K + q(L_{O1} + L_{O2} - 2L_{O6})]\}$$
$$+C_{12}\{8C_{44}(2K - 3L_2 + qL_{O2}) + 3C_{12}[8K - 8L_2 + q(L_{O1} + 3L_{O2}$$
$$-2L_{O6})]\} + 3C_{11}^2[8K + 8L_2 + q(3L_{O1} + L_{O2} - 6L_{O6})]\},$$
$$g_{s1}^{b1} = -4C_{11}[C_{44}(6K + 3L_1 - 2L_2 + 3qL_{O1} + 3qL_{O3}) + 3C_{12}(6K + L_1 - L_2 + 2L_3 \tag{36}$$
$$+2qL_{O1} + 3qL_{O2} + qL_{O3} - 4qL_{O4} + 2qL_{O6})] + C_{12}[4C_{44}(6K + 3L_1 - 8L_2 + 3qL_{O1}$$
$$+3qL_{O3}) + 3C_{12}(12K + 4L_1 - 12L_2 + 4L_3 + 5qL_{O1} + 3qL_{O2} + 4qL_{O3} - 8qL_{O4}$$
$$+4qL_{O6})] + 3C_{11}^2\{12K + 8L_2 + 3[2L_3 + q(L_{O1} + 3L_{O2} - 4L_{O4} + 2L_{O6})]\},$$

$$g_{c2}^{b1} = -6\{C_{12}\{-8C_{44}(2K - 3L_2 + qL_{O2}) + C_{12}[8K + q(3L_{O1} + L_{O2} - 6L_{O6})]\}$$
$$-4C_{11}\{2C_{44}(-2K + L_2 - qL_{O2}) + C_{12}[4K + q(L_{O1} + L_{O2} - 2L_{O6})]\}$$
$$+C_{11}^2[8K + q(L_{O1} + 3L_{O2} - 2L_{O6})]\},$$
$$g_{s2}^{b1} = \sqrt{3}\{C_{11}^2[12K + 4L_1 + 2L_3 + q(5L_{O1} + 3L_{O2} + 4L_{O3} - 4L_{O4} + 2L_{O6})] - 4C_{11}[-C_{44} \tag{37}$$
$$\times (6K + 3L_1 - 2L_2 + 3qL_{O1} + 3qL_{O3}) + C_{12}(6K + L_1 - L_2 + 2L_3 + 2qL_{O1} + 3qL_{O2}$$
$$+qL_{O3} - 4qL_{O4} + 2qL_{O6})] + C_{12}\{-4C_{44}(6K + 3L_1 - 8L_2 + 3qL_{O1} + 3qL_{O3})$$
$$+C_{12}\{12K - 4L_2 + 3[2L_3 + q(L_{O1} + 3L_{O2} - 4L_{O4} + 2L_{O6})]\}\}\},$$

and $C_k = 3C_{11}^2 + 10C_{11}C_{44} - 3C_{12}(C_{12} + 2C_{44})$.

The solution of periodic displacement field in the skyrmion phase is composed of three types of triple-Q structures, defined by

$$u_i^{S1}(\mathbf{r}) = \mathbf{Re}\big[u_i^{a1}e^{i\mathbf{q}_1\mathbf{r}} + u_i^{b1}e^{i\mathbf{q}_2\mathbf{r}} + u_i^{c1}e^{i\mathbf{q}_3\mathbf{r}}\big],$$
$$u_i^{S2}(\mathbf{r}) = \mathbf{Re}\big[u_i^{a2}e^{2i\mathbf{q}_1\mathbf{r}} + u_i^{b2}e^{2i\mathbf{q}_2\mathbf{r}} + u_i^{c2}e^{2i\mathbf{q}_3\mathbf{r}}\big], \tag{38}$$
$$u_i^{S3}(\mathbf{r}) = \mathbf{Re}\big[u_i^{ab1}e^{i(\mathbf{q}_1-\mathbf{q}_2)\mathbf{r}} + u_i^{ac1}e^{i(\mathbf{q}_1-\mathbf{q}_3)\mathbf{r}} + u_i^{bc1}e^{i(\mathbf{q}_2-\mathbf{q}_3)\mathbf{r}}\big].$$

With eq. (38), eq. (24) can be recasted as $\mathbf{u} = \mathbf{u}^c + \mathbf{u}^{S1}(\mathbf{r}) + \mathbf{u}^{S2}(\mathbf{r}) + \mathbf{u}^{S3}(\mathbf{r})$.

Eqs. (25-37) show that the magnitude of the periodic displacement field depends on the magnetization as well as the magnetoelastic effects. To be more specific, in terms of the magnetization, $u_i^{S1}(\mathbf{r})$ depends on both $m_3$ and $m_q$ defined in eq. (23), while $u_i^{S2}(\mathbf{r})$ and $u_i^{S3}(\mathbf{r})$ depend merely on $m_q$; in terms of the magnetoelastic effects, $u_i^{S1}(\mathbf{r})$ and $u_i^{S3}(\mathbf{r})$ depend on $K$, which represents the dominant magnetoelastic term, while $u_i^{S2}(\mathbf{r})$ is independent of $K$.

It is easy to verify that in the skyrmion phase, the eigenstrains $\varepsilon_{ij}^*$ do not satisfy the compatibility condition. Similar to explanation given in section 3, the part of eigenstrains that doesn't satisfy the compatibility will generate periodic elastic stresses. The components of the three triple-Q stress tensors can be obtained as

$$\sigma_{11}^{Si} = C_{11}e_{11}^{Si} + C_{12}(e_{22}^{Si} + e_{33}^{Si}),$$
$$\sigma_{22}^{Si} = C_{11}e_{22}^{Si} + C_{12}(e_{11}^{Si} + e_{33}^{Si}),$$
$$\sigma_{33}^{Si} = C_{11}e_{33}^{Si} + C_{12}(e_{11}^{Si} + e_{22}^{Si}),$$
$$\sigma_{23}^{Si} = 2C_{44}e_{23}^{Si},$$
$$\sigma_{13}^{Si} = 2C_{44}e_{13}^{Si},$$
$$\sigma_{12}^{Si} = 2C_{44}e_{12}^{Si}, \quad \text{(for } i = 1,2,3\text{)} \tag{39}$$

where

$$e_{ij}^{S1} = \frac{1}{2}(u_{i,j}^{S1} + u_{j,i}^{S1}) - \mathbf{Re}[\varepsilon_{ij}^{a1*}e^{i\mathbf{q}_1\mathbf{r}} + \varepsilon_{ij}^{b1*}e^{i\mathbf{q}_2\mathbf{r}} + \varepsilon_{ij}^{c1*}e^{i\mathbf{q}_3\mathbf{r}}],$$
$$e_{ij}^{S2} = \frac{1}{2}(u_{i,j}^{S2} + u_{j,i}^{S2}) - \mathbf{Re}[\varepsilon_{ij}^{a2*}e^{2i\mathbf{q}_1\mathbf{r}} + \varepsilon_{ij}^{b2*}e^{2i\mathbf{q}_2\mathbf{r}} + \varepsilon_{ij}^{c2*}e^{2i\mathbf{q}_3\mathbf{r}}], \tag{40}$$
$$e_{ij}^{S3} = \frac{1}{2}(u_{i,j}^{S3} + u_{j,i}^{S3}) - \mathbf{Re}[\varepsilon_{ij}^{ab1*}e^{i(\mathbf{q}_1-\mathbf{q}_2)\mathbf{r}} + \varepsilon_{ij}^{ac1*}e^{i(\mathbf{q}_1-\mathbf{q}_3)\mathbf{r}} + \varepsilon_{ij}^{bc1*}e^{i(\mathbf{q}_2-\mathbf{q}_3)\mathbf{r}}].$$

The analytical expressions of $\sigma_{kl}^{Si}, (i,k,l = 1,2,3)$ can be obtained by substituting eqs. (25-38) into eqs. (39, 40), which are too lengthy to be expanded here. This result indicates that in bulk materials free from any mechanical loads, appearance of magnet skyrmions is always accompanied by nontrivial periodic stress fields.

**Field configuration of the triple-Q elastic structures for MnSi and discussion**

For $MnSi$, the configuration of the three triple-Q displacement structures are plotted in Figure 1(a-c) using the magnetization obtained through free-energy minimization at temperature 4K and zero magnetic field using the extended micromagnetic model[15]. From Figure 1(a-c), we learn that $\mathbf{u}^{S1}(\mathbf{r})$, $\mathbf{u}^{S2}(\mathbf{r})$ and $\mathbf{u}^{S3}(\mathbf{r})$ have field configurations that are different from each other and also different from the magnetic skyrmions. However, they all form hexagonal networks of localized fields, while all the three triple-Q displacement structures have only in-plane components. This can be explained from eqs. (25-34), where all the displacement components in $z-$axis rely only on $L_{O4}, L_{O5}$, and $L_{O6}$, which are set to be zero for $MnSi$. In fact, since $L_{O4}, L_{O5}$, and $L_{O6}$ represent high order magnetoelastic effects[15], the smallness of $z-$component of the skyrmion induced displacement field is guaranteed for any B20 compound.

From Figure 1(d), we see the total periodic displacement field denoted by $\mathbf{u}^{S1}(\mathbf{r}) + \mathbf{u}^{S2}(\mathbf{r}) + \mathbf{u}^{S3}(\mathbf{r})$ appears to have the same periodicity with $\mathbf{u}^{S1}(\mathbf{r})$, which can be explained as follow. The lattice vectors of the periodic displacement structure $\mathbf{u}^{S1}(\mathbf{r})$, denoted by $\mathbf{a}_1$ and $\mathbf{a}_2$, satisfy

$$\begin{aligned}\mathbf{a}_1 \cdot \mathbf{q}_1 &= 1, \\ \mathbf{a}_1 \cdot \mathbf{q}_2 &= 0, \\ \mathbf{a}_2 \cdot \mathbf{q}_1 &= 0, \\ \mathbf{a}_2 \cdot \mathbf{q}_2 &= 1.\end{aligned} \tag{41}$$

The periodicity of $u_i^{S1}(\mathbf{r})$ can generally be described by $u_i^{S1}(\mathbf{r} + 2n_1\pi\mathbf{a}_1 + 2n_2\pi\mathbf{a}_2) = u_i^{S1}(\mathbf{r})$, where $n_1$ and $n_2$ are arbitrary integers. From eq. (38), we can easily prove that $u_i^{S2}(\mathbf{r} + 2n_1\pi\mathbf{a}_1 + 2n_2\pi\mathbf{a}_2) = u_i^{S3}(\mathbf{r})$ and $u_i^{S3}(\mathbf{r} + 2n_1\pi\mathbf{a}_1 + 2n_2\pi\mathbf{a}_2) = u_i^{S3}(\mathbf{r})$. Hence $u_i^{S1}(\mathbf{r}) + u_i^{S2}(\mathbf{r}) + u_i^{S3}(\mathbf{r})$ and $u_i^{S1}(\mathbf{r})$ (or $\mathbf{M}(\mathbf{r})$) shares the same lattice vectors and period. For bulk chiral magnets, this period, determined by the strength of the DM interaction and the stiffness of the exchange energy density, is independent of the period of the underlying atomic lattice. This explains directly why magnetic skyrmion materials are incommensurate systems, and also provides quantitative description of the deformed lattice configuration of the skyrmion materials.

The maximum achievable displacement of $\mathbf{u}^{S1}$, $\mathbf{u}^{S2}$, $\mathbf{u}^{S3}$ and $\mathbf{u}^{S1} + \mathbf{u}^{S2} + \mathbf{u}^{S3}$ are listed in Table 1, from which we can see $\max_{\mathbf{r}}|\mathbf{u}^{S3}| > \max_{\mathbf{r}}|\mathbf{u}^{S1}| \gg \max_{\mathbf{r}}|\mathbf{u}^{S2}|$. The magnitude of $\mathbf{u}^{S1}$ and $\mathbf{u}^{S3}$ is significantly larger than that of $\mathbf{u}^{S2}$, since the solution of $\mathbf{u}^{S1}$ and $\mathbf{u}^{S3}$ is related to the exchange-interaction-induced magnetoelastic coupling term with coefficient $K$, which is the dominant term among all magnetoelastic coupling effects (for $MnSi$, the magnitude of $K$ is at least an order of magnitude larger than the coefficient of other magnetoelastic terms[15]).

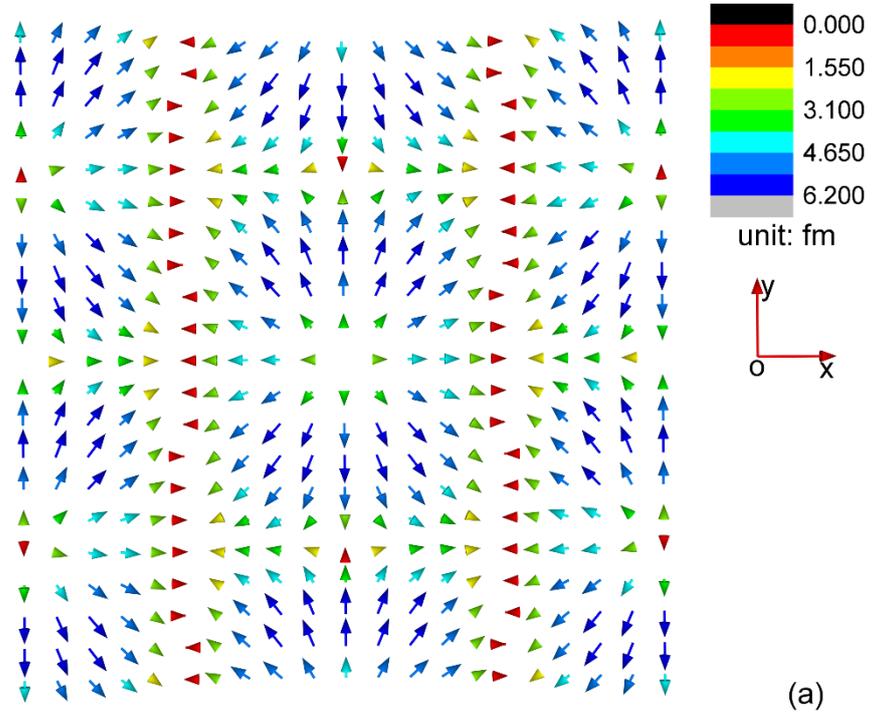

(a)

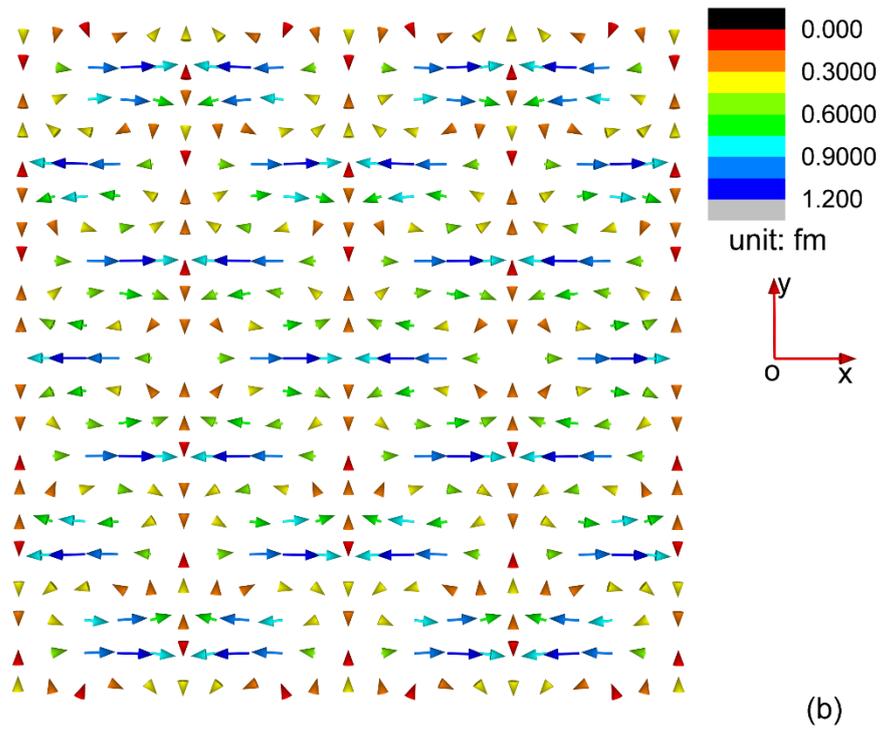

(b)

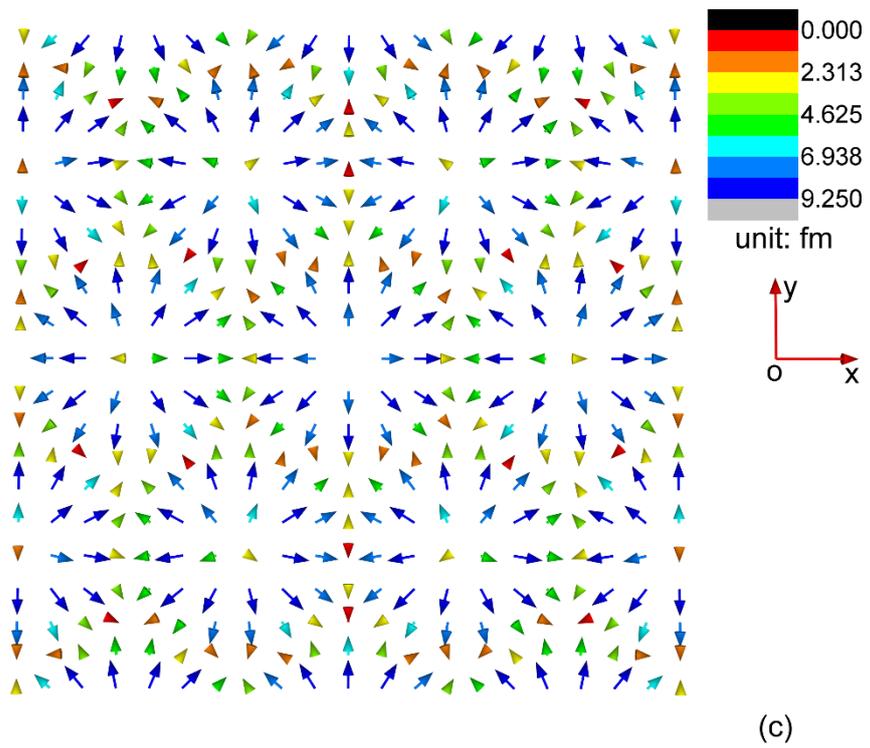

(c)

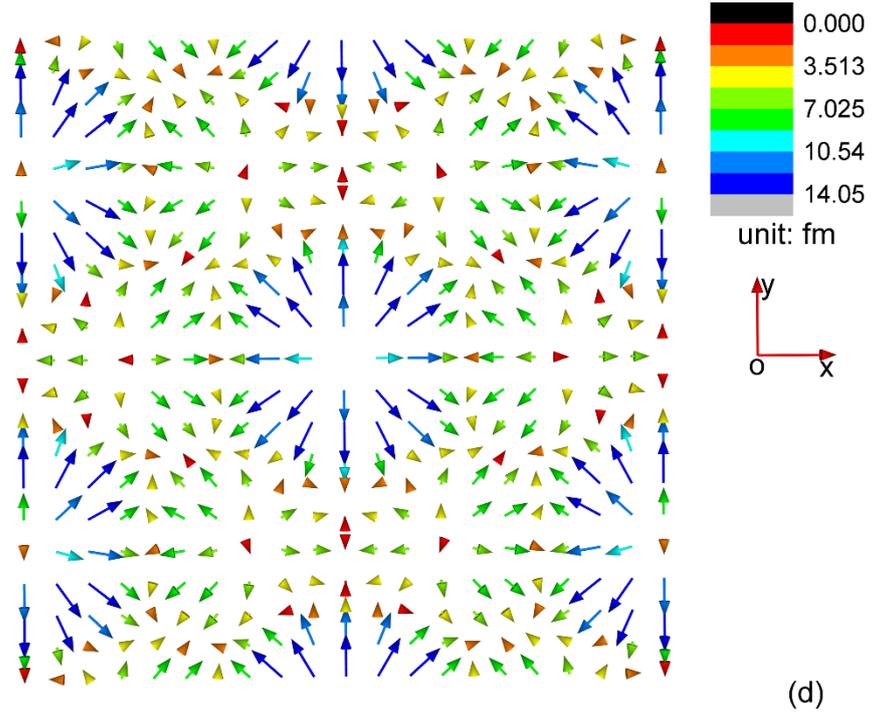

(d)

Figure 1. Configuration of the three triple-Q displacement structures at temperature 4K and magnetic field 0T: (a)$\mathbf{u}^{S1}$, (b)$\mathbf{u}^{S2}$, (c)$\mathbf{u}^{S3}$, and (d)$\mathbf{u}^{S1} + \mathbf{u}^{S2} + \mathbf{u}^{S3}$. The arrows represent the direction of the displacement vectors which vary in space, and the length of the arrows reflects the magnitude of the displacement vectors.

Table 1. Magnitude of maximum achievable displacement for the three triple-Q displacement structures of $MnSi$ and their sum calculated at temperature 4K and magnetic field 0T

| Triple-Q displacement structures: | $\mathbf{u}^{S1}$ | $\mathbf{u}^{S2}$ | $\mathbf{u}^{S3}$ | $\mathbf{u}^{S1} + \mathbf{u}^{S2} + \mathbf{u}^{S3}$ |
|---|---|---|---|---|
| Maximum displacement (nm): | $6.20 \times 10^{-6}$ | $1.20 \times 10^{-6}$ | $9.25 \times 10^{-6}$ | $1.41 \times 10^{-5}$ |
| Magnitude of wave vector: | $q$ | $2q$ | $\sqrt{3}q$ | $q$ |

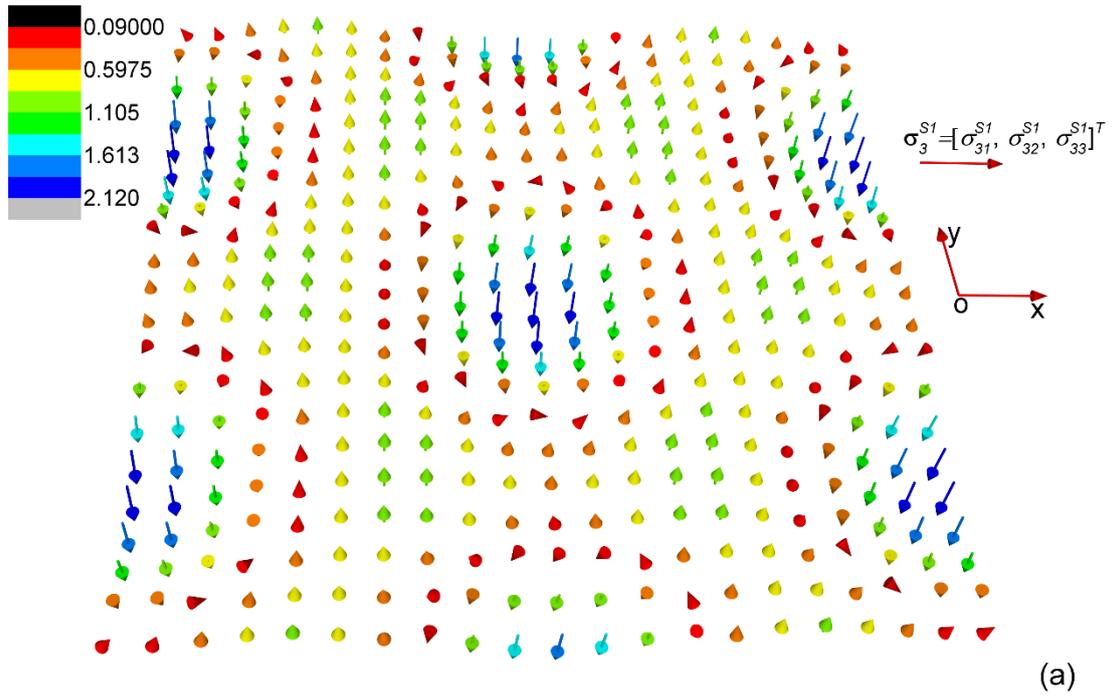

(a)

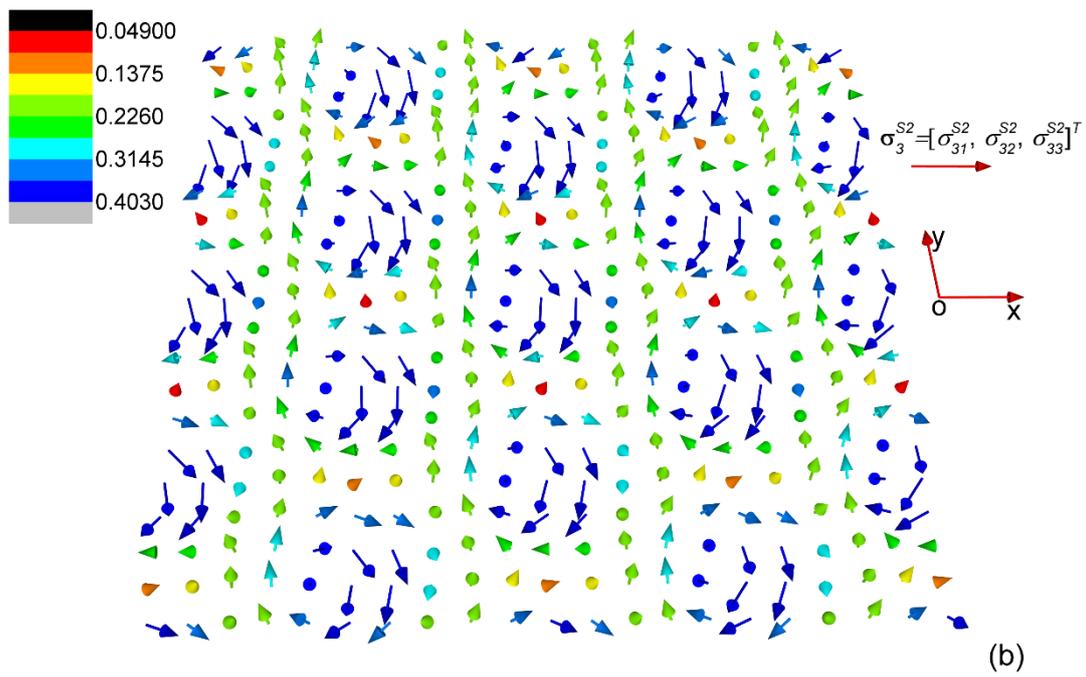

(b)

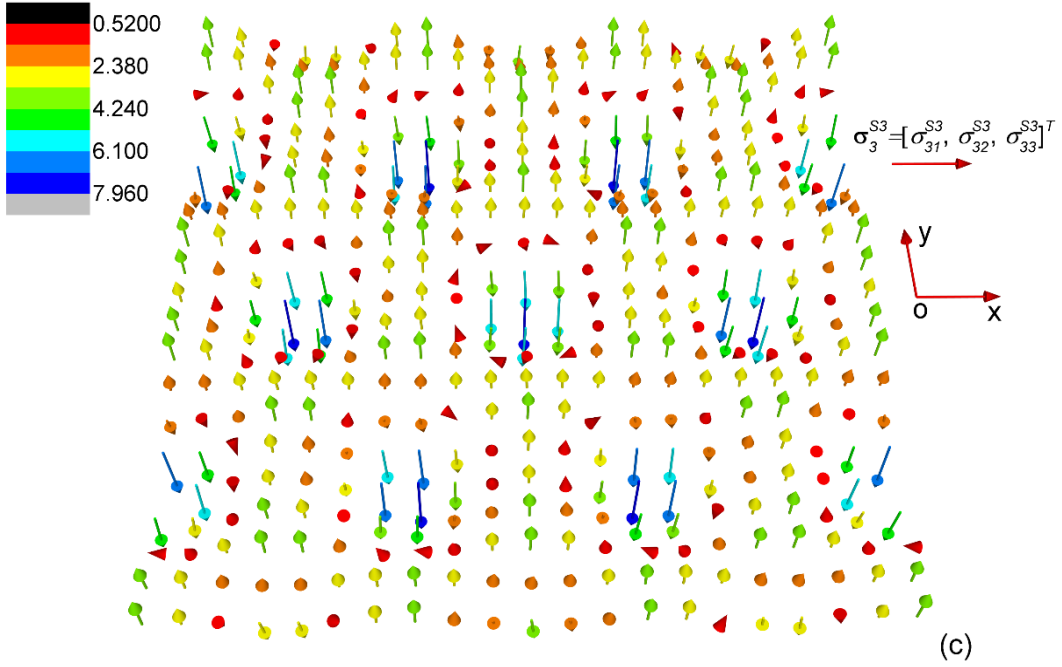

Figure 2. Configuration of the components of the three triple-Q stress tensors defined on the (001) plane (or the $xy$ plane) at temperature 4K and magnetic field 0T: (a) $\sigma_{3i}^{S1}$, (b) $\sigma_{3i}^{S2}$, and (c) $\sigma_{3i}^{S3}$. The unit used in all three figures is MPa.

When solving the stress field of $MnSi$, it is found that the stress components defined on (001) plane $\sigma_{3l}^{Si}, (i, l = 1, 2, 3)$ is generally much larger than other stress components. This can be understood since the periodic in-plane displacements release most of the eigenstrains defined on (100) plane and (010) plane while the other incompatible eigenstrains cause stresses mainly on the (001) plane. The field configuration of $\boldsymbol{\sigma}_3^{Si} = [\sigma_{31}^{Si}, \sigma_{32}^{Si}, \sigma_{33}^{Si}]^T, (i = 1, 2, 3)$ is plotted in Figure 2(a-c), where we can see that $\sigma_{33}^{Si}, (i = 1, 2, 3)$ is the most significant stress component.

*Variation of the field configuration with temperature and magnetic field*

The periodic elastic fields solved as functions of the equilibrium magnetization, should vary with temperature and magnetic field. When the temperature is increased from 0K to the critical temperature, it is found that the field configuration of all the triple-Q structures is merely changed, but the magnitude of vectors gradually decreases to zero. This is understood since the components of the magnetization gradually decrease to zero as the temperature approaches the critical temperature.

On the other hand, we find that the field configuration of $\mathbf{u}^{S1}$ and $\boldsymbol{\sigma}_3^{S1}$ is very sensitive to variation of the magnitude of applied magnetic field while the field configuration of other triple-Q field structures is not. In Figure 3(a-d) and Figure 4(a-d), we plot the variation of the field configuration of $\mathbf{u}^{S1}$ and $\boldsymbol{\sigma}_3^{S1}$ when the magnetic field gradually increases from 0.1T to 0.4T. We find an interesting phenomenon that the field configuration of both $\mathbf{u}^{S1}$ and $\boldsymbol{\sigma}_3^{S1}$ undergoes a "configurational reversal" when the external magnetic field increases: comparing Figure 3(a) and Figure 3(d) (Figure 4(a) and Figure 4(d)), it is observed that field configuration of $\mathbf{u}^{S1}$ ($\boldsymbol{\sigma}_3^{S1}$) plotted at applied field 0.1T and 0.4T shares similar pattern but with opposite direction of all the field vectors. One should notice that such a reversal of elastic field configuration does not affect the magnetic state of the material (i.e., no magnetic phase transition occurs, and the magnetization is not reversed). Through thermodynamic analysis within the mean-field theory[15], we already known that the variation of $m_3$ and $m_q$ with external magnetic field in the skyrmion phase is insignificant. $\mathbf{u}^{S2}$, $\mathbf{u}^{S3}$, $\boldsymbol{\sigma}_3^{S2}$, and $\boldsymbol{\sigma}_3^{S3}$ depend merely on $m_q$, and so their field configuration merely changes with external magnetic field. To understand why $\mathbf{u}^{S1}$ and $\boldsymbol{\sigma}_3^{S1}$ are sensitive to external magnetic field, we first consider the $K$–dependent terms in $\mathbf{u}^{S1}$ and $\boldsymbol{\sigma}_3^{S1}$. It is found after manipulation that the $\mathbf{u}^{S1}$ and $\boldsymbol{\sigma}_3^{S1}$ depend on $K$ linearly, with a coefficient proportional to $4\sqrt{3}m_3 - 3m_q$, which vanishes at $\frac{m_q}{m_3} \approx 2.31$. For $MnSi$, as the magnetic field increases from 0T to 0.4T, the equilibrium value of $\frac{m_q}{m_3}$ in the skyrmion phase decreases from 2.59 to 2.15. As a result, the coefficient of the linear term of $K$ changes sign as the magnetic field increases, which is responsible for the configurational reversal of $\mathbf{u}^{S1}$ and $\boldsymbol{\sigma}_3^{S1}$.

In Figure 5(a-d), we plot the variation of the field configuration of $\mathbf{u}^{S1} + \mathbf{u}^{S2} + \mathbf{u}^{S3}$ when the magnetic field gradually increases from 0.1T to 0.4T. We find that the configurational reversal of $\mathbf{u}^{S1}$ is smeared by $\mathbf{u}^{S2} + \mathbf{u}^{S3}$, which is almost unchanged for increasing magnetic field. Yet, by comparing Figure 5(a) and Figure 5(d) we see that the location of a "significant outburst" changes from the center in Figure 5(a) to six adjacent points of the center in Figure 5(d).

In eq. (23), the magnetization function introduced is specified for the case where the external magnetic field is applied along the z-axis. Concerning the magnetoelastic effects described in eq. (2), all terms except $KM^2\varepsilon_{ii}$ are anisotropic, which means that the solution of displacement field, as well as the field configurations, all vary with the direction of applied field. In other words, the solution given in eqs. (25-39) are specified for the case where the external field is applied along the z-axis.

### *Condition for occurrence of a configurational reversal of $\mathbf{u}^{S1}$ and $\boldsymbol{\sigma}_3^{S1}$ for B20 compounds*

The analytical solution of $\mathbf{u}^{S1}$ and $\boldsymbol{\sigma}_3^{S1}$ derived in eqs. (38, 39) applies to any B20 compounds. As a result, the coefficient of the linear term of $K$ in $\mathbf{u}^{S1}$ and $\boldsymbol{\sigma}_3^{S1}$ always vanishes at $\frac{m_q}{m_3} \approx 2.31$. The occurrence of a configurational reversal of $\mathbf{u}^{S1}$ and $\boldsymbol{\sigma}_3^{S1}$ is garanteed if the equilibrium value of $\frac{m_q}{m_3}$ in the skyrmion phase at 0T is larger than 2.31 and the equilibrium value of $\frac{m_q}{m_3}$ in the skyrmion phase at some critical magnetic field is smaller than 2.31. After analysing the free-energy minimization process within the extended micromagnetic model[15], we find that the equilibrium value of $\frac{m_q}{m_3}$ is determined by the Landau expansion terms $w_L = \alpha(T - T_0)\mathbf{M}^2 + \beta\mathbf{M}^4$ and the Dzyaloshinskii-Moriya (DM) coupling term $b\mathbf{M} \cdot (\nabla \times \mathbf{M})$, where $T$ denotes the temperature. Neglecting the DM coupling term and substituting eq. (23) into $w_L$, we find that the equilibrium value of $\frac{m_q}{m_3}$ in the skyrmion phase equals to 2.23, regardless of the value of $\alpha$, $\beta$ and $T$. Adding the DM coupling term back to the functional, it can be calculated that once $\frac{bq}{|\alpha(T-T_0)|} > 0.08$, we always have $\frac{m_q}{m_3} > 2.31$ in the skyrmion phase, where $q$ denotes the wave number. Hence the stronger the DM coupling, the larger $\frac{m_q}{m_3}$. For $MnSi$, $\frac{bq}{|\alpha(T-T_0)|} \approx 0.26$ at 0K. On the other hand, application of external magnetic field along the z-axis will inevitably increase $m_3$ and decrease $m_q$, which leads to a decrease of $\frac{m_q}{m_3}$. Concerning the above analysis, we think the occurrence of a configurational reversal of $\mathbf{u}^{S1}$ and $\boldsymbol{\sigma}_3^{S1}$ with increasing magnetic field is more of a general phenomenon for any B20 compound than a specific issue for $MnSi$. Even if a configurational reversal does not occur, the variation of $\mathbf{u}^{S1}$ and $\boldsymbol{\sigma}_3^{S1}$ with external magnetic field should be significant.

*Technological interest of elastic triple-Q structures in chiral magnets*

Magnetic skyrmions are regarded as one type of possible information carrier, since their motion in materials can be manipulated by small current density[6,7,20], and their existence can be manipulated by various approaches[21-23]. Stem from the intrinsic magnetoelastic coupling in chiral magnets, the elastic triple-Q structures always move together with the magnetic skyrmions. This provides the possibility of identifying the existence of magnetic skyrmions by checking the localized elastic state. Moreover, the field configuration of the elastic triple-Q structures having the same periodicity with the skyrmions is sensitive to external magnetic field, which means that we can have opposite elastic state for almost unchanged magnetic state. From the solution obtained in eqs. (25-37), the magnitude of the periodic elastic field can be enhanced by increasing the magnetoelastic coefficients and also the size (the wavelength) of an individual skyrmion, which may be achieved by choosing appropriate materials.

Besides data-storage devices, magnetic skyrmions are also promising for developing new microwave applications, since they excite gigahertz collective spin vibration modes when exposed to magnetic microwave [14,24,25]. Existence of the elastic triple-Q structures provides a variety of options for development of related technology, such as novel magneto-acoustic actuators and sensors. For this purpose, it is of interest to derive the corresponding collective elastic field vibration modes from the known collective spin vibration modes and study their coupling in dynamical conditions.

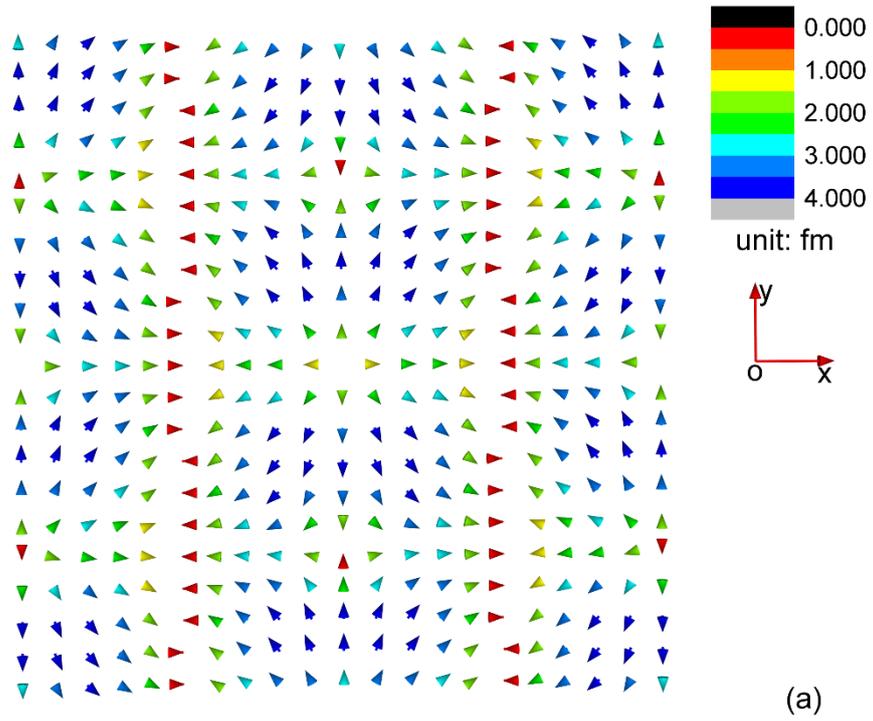

(a)

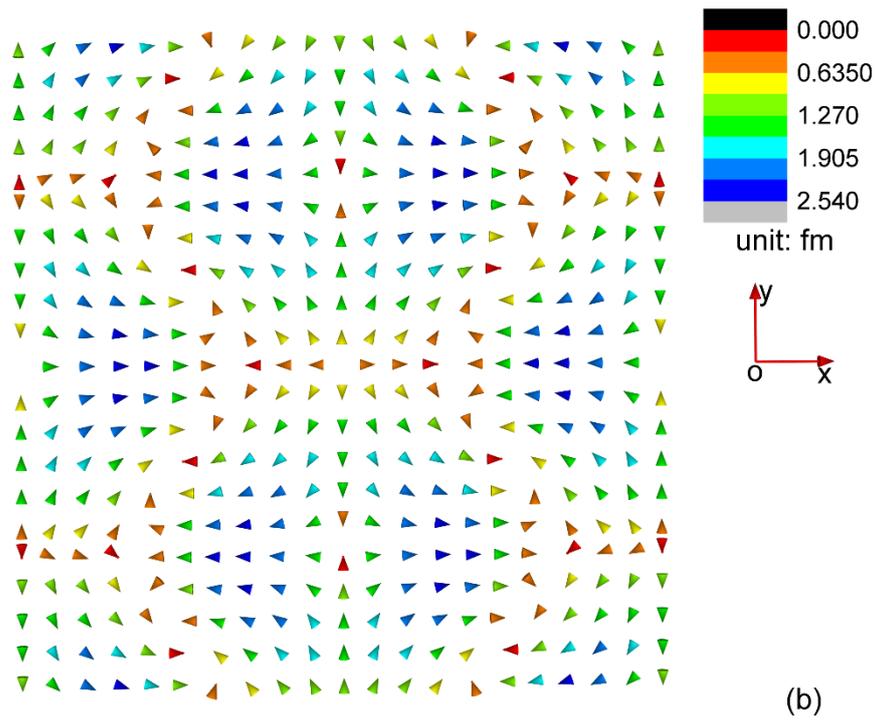

(b)

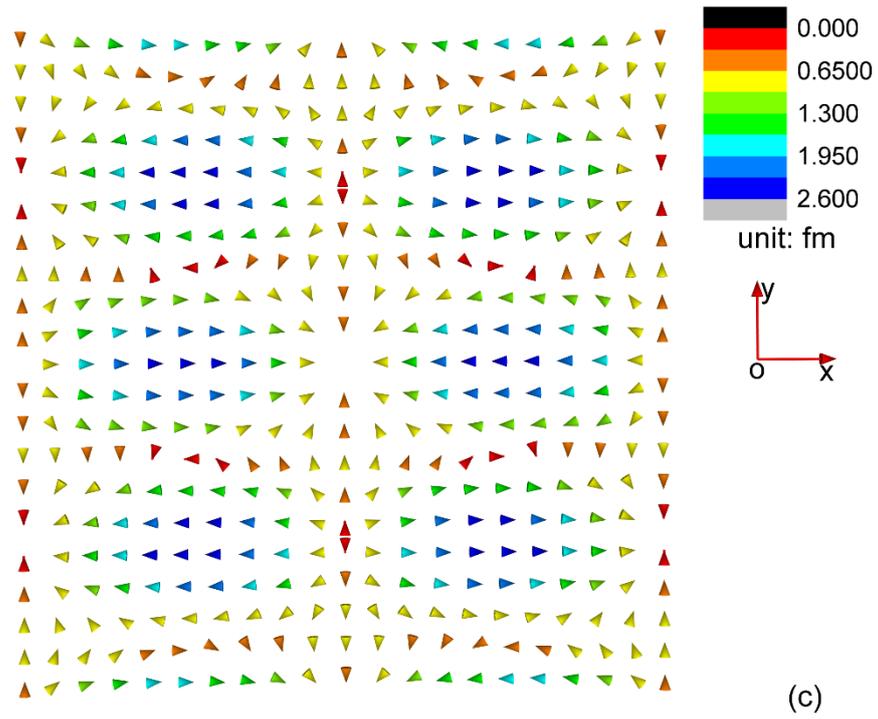

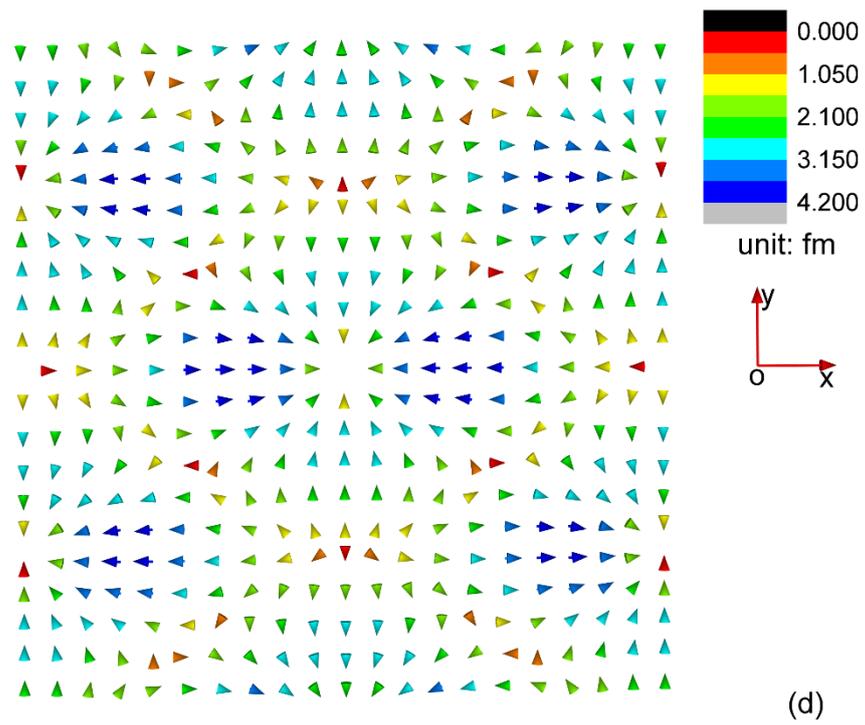

Figure 3. Configuration of $\mathbf{u}^{S1}$ at temperature 4K and magnetic field (a)0.1T, (b)0.2T, (c)0.3T, and (d)0.4T.

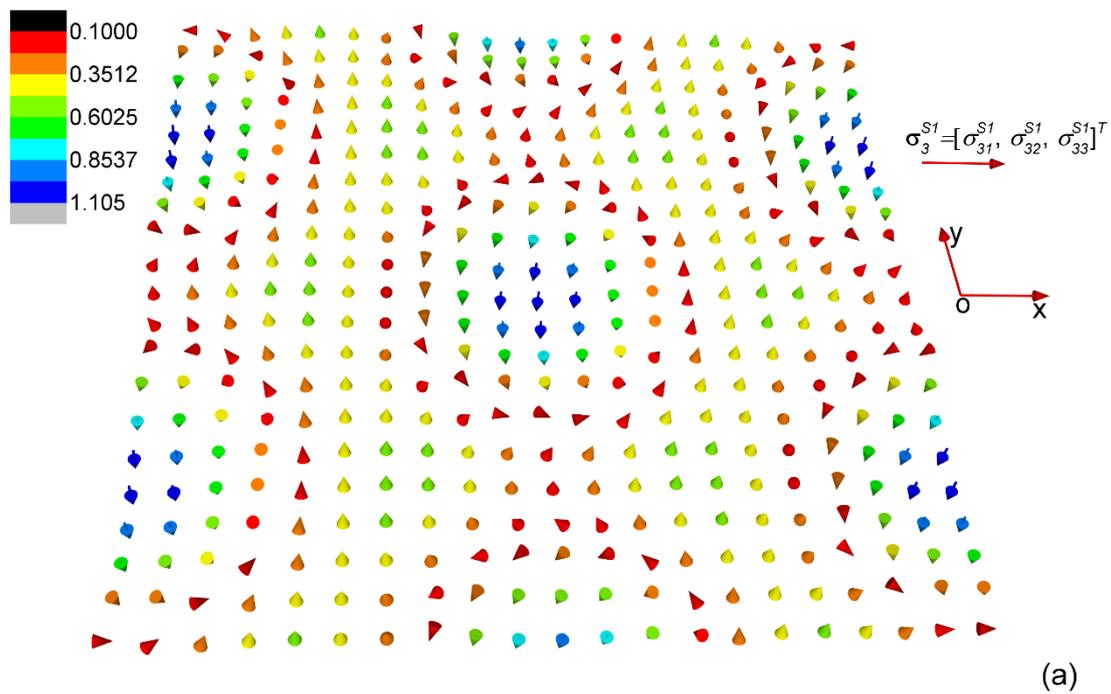

(a)

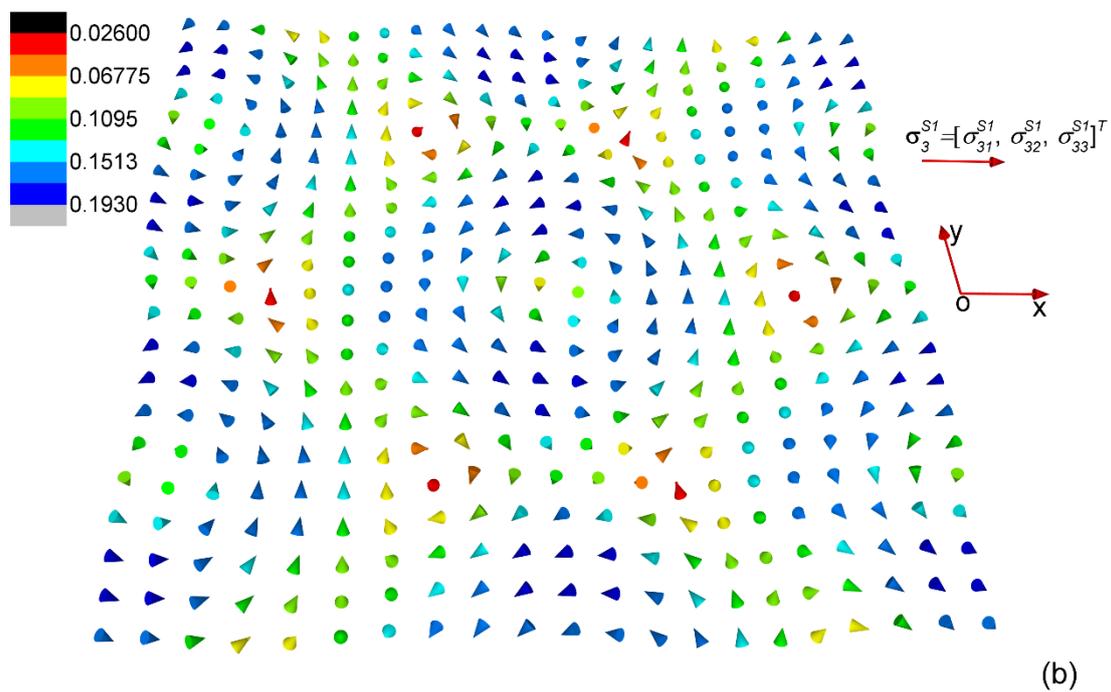

(b)

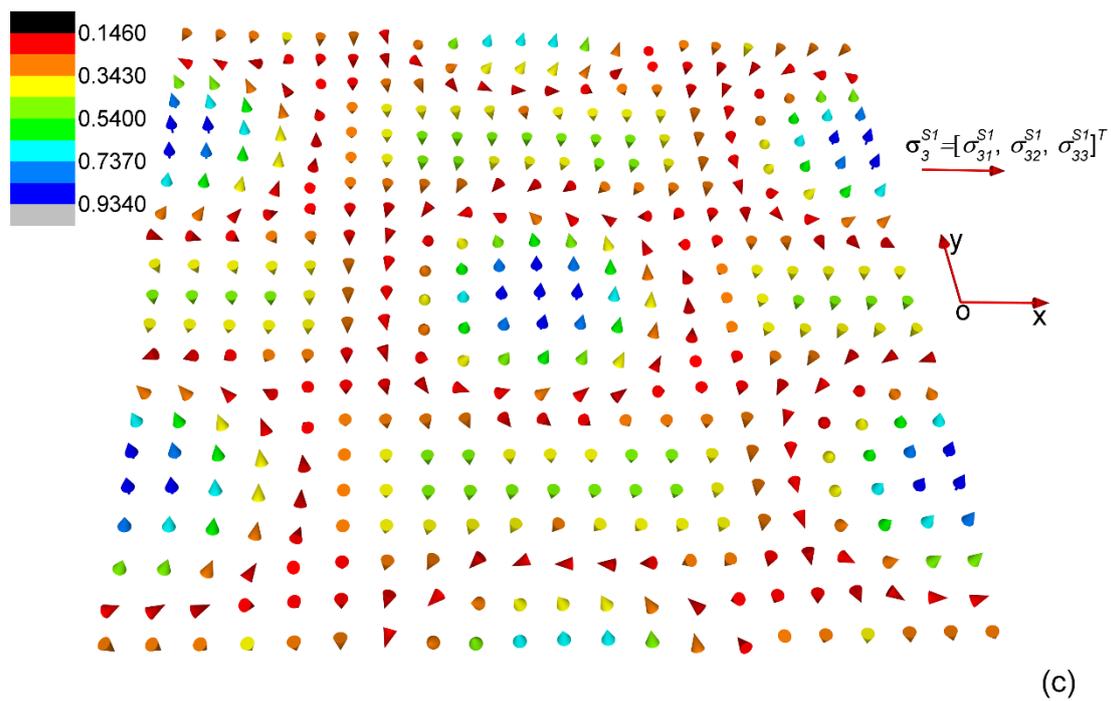
(c)

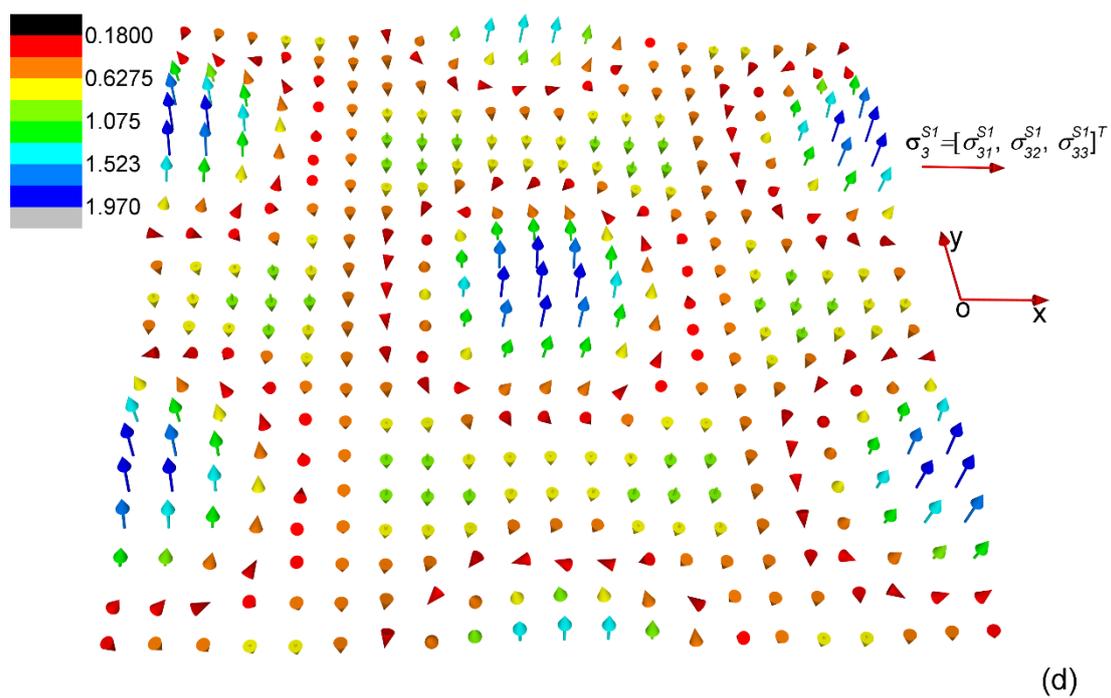
(d)

Figure 4. Configuration of $\boldsymbol{\sigma}_3^{S1}$ defined on the (001) plane (or the $xy$ plane) at temperature 4K and magnetic field (a)0.1T, (b)0.2T, (c)0.3T, and (d)0.4T.

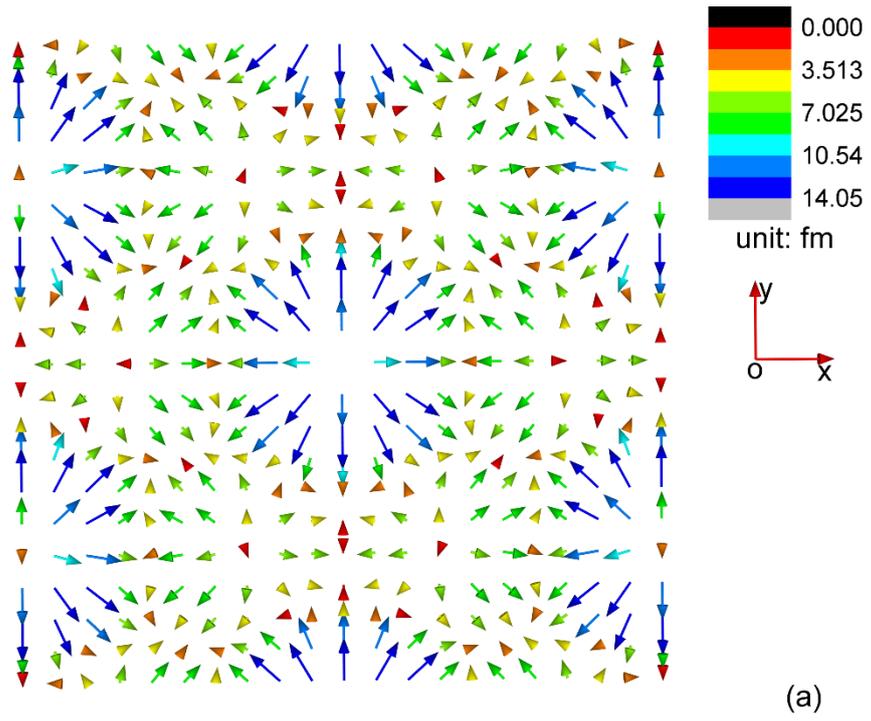

(a)

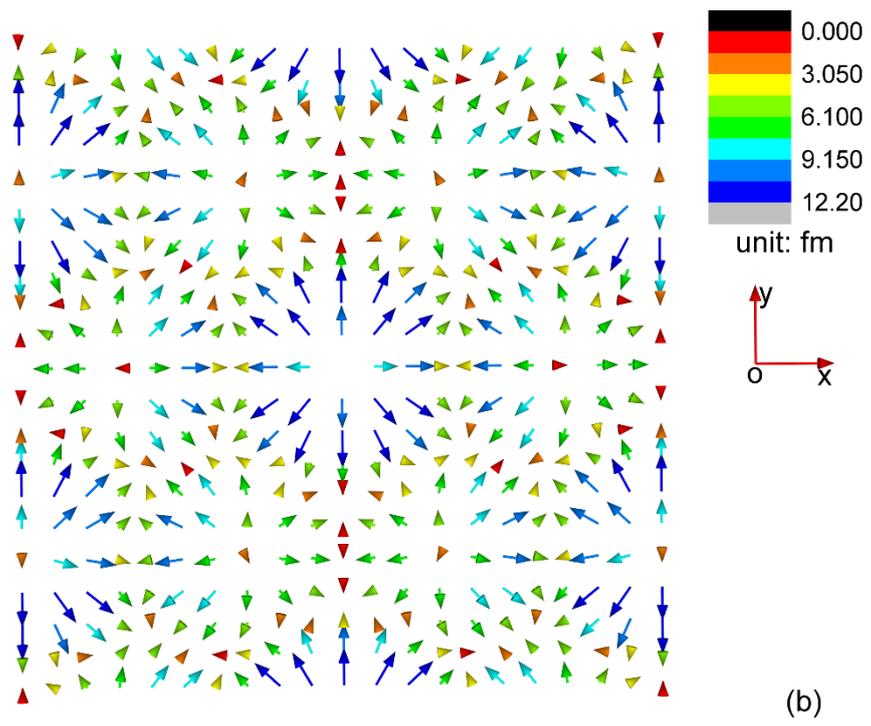

(b)

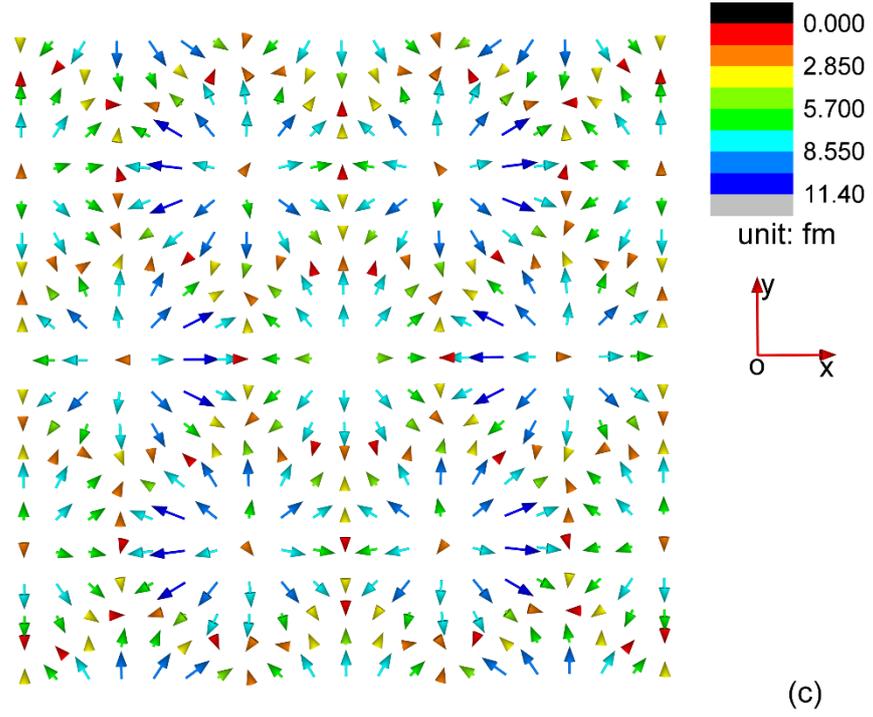

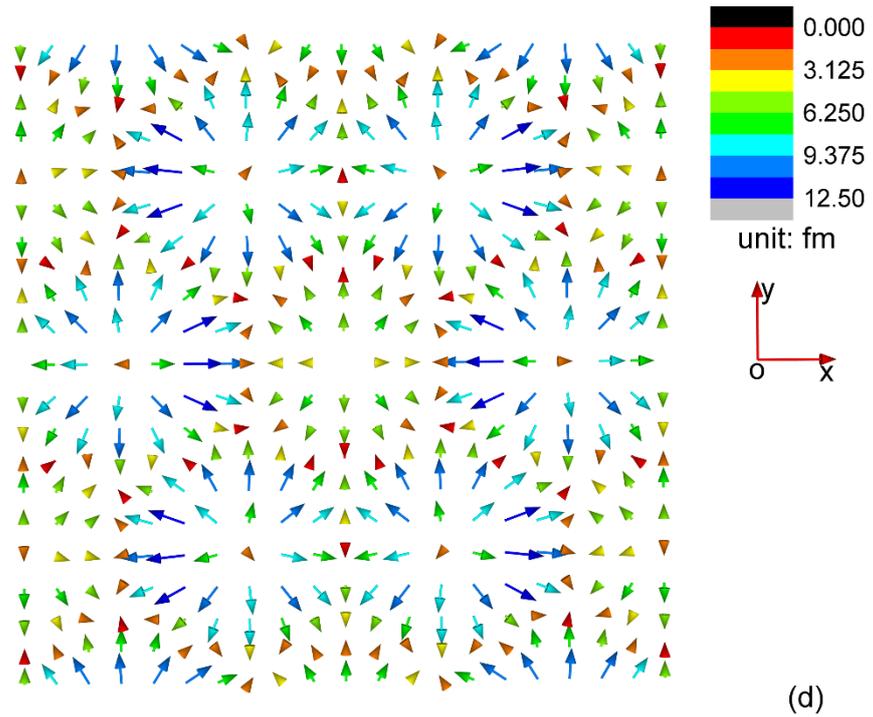

Figure 5. Configuration of $\mathbf{u}^{S1} + \mathbf{u}^{S2} + \mathbf{u}^{S3}$ at temperature 4K and magnetic field (a)0.1T, (b)0.2T, (c)0.3T, and (d)0.4T.

## Conclusion

We obtain the analytical solution of the periodic elastic fields for the eigenstrain problem in chiral magnets. In the skyrmion phase, the nonlinear magnetoelastic coupling leads to formation of three types of triple-Q elastic field structures. For MnSi, the triple-Q displacement structure $\mathbf{u}^{S1}$ and the triple-Q stress structure $\boldsymbol{\sigma}_3^{S1}$ are found to undergo a configurational reversal when the magnetic field increases from 0T to 0.4T. Through thermodynamic analysis, we find that such a configurational reversal is likely to occur for any B20 compound. It will be interesting to experimentally detect the periodic elastic structure, and further discuss the intrinsic vibration modes of these elastic triple-Q structures.

**Acknowledgement**: We would like to thank Xuejin Wan for helpful discussion. The work was supported by the NSFC (National Natural Science Foundation of China) through the fund 11302267.

**Author contributions**:
Y. Hu conceived the idea and finished the analytical deduction. Y. Hu and B. Wang discussed the results for revision. Y. Hu and B. Wang co-wrote the manuscript.

**Competing financial interests:** The authors declare no competing financial interests.

**References**
1   Muhlbauer, S. *et al.* Skyrmion Lattice in a Chiral Magnet. *Science* 323, 915-919, doi:DOI 10.1126/science.1166767 (2009).
2   Yu, X. Z. *et al.* Real-space observation of a two-dimensional skyrmion crystal. *Nature* 465, 901-904, doi:10.1038/nature09124 (2010).
3   Yu, X. Z. *et al.* Near room-temperature formation of a skyrmion crystal in thin-films of the helimagnet FeGe. *Nat Mater* 10, 106-109, doi:10.1038/nmat2916 (2011).
4   Dzyaloshinskii, I. Theory of helicoidal structures in antiferromagnets. 1. Nonmetals. *Sov. Phys. JETP* 19, 960-971 (1964).
5   Bak, P. & Jensen, M. H. Theory of helical magnetic structures and phase transitions in MnSi and FeGe. *Journal of Physics C: Solid State Physics* 13, L881 (1980).
6   Schulz, T. *et al.* Emergent electrodynamics of skyrmions in a chiral magnet. *Nat Phys* 8, 301-304, doi:Doi 10.1038/Nphys2231 (2012).
7   Fert, A., Cros, V. & Sampaio, J. Skyrmions on the track. *Nat Nanotechnol* 8, 152-156 (2013).
8   Neubauer, A. *et al.* Topological Hall effect in the A phase of MnSi. *Phys Rev Lett* 102, 186602, doi:Artn 186602 Doi 10.1103/Physrevlett.102.186602 (2009).
9   Drozdov, I. K. *et al.* Extra Spin-Wave Mode in Quantum Hall Systems: Beyond the Skyrmion Limit. *Phys Rev Lett* 104, doi:Artn 136804 Doi 10.1103/Physrevlett.104.136804 (2010).
10  Butenko, A. B., Leonov, A. A., Rossler, U. K. & Bogdanov, A. N. Stabilization of skyrmion textures by uniaxial distortions in noncentrosymmetric cubic helimagnets. *Phys Rev B* 82, doi:Artn 052403 Doi 10.1103/Physrevb.82.052403 (2010).
11  Huang, S. X. & Chien, C. L. Extended Skyrmion phase in epitaxial FeGe(111) thin films. *Phys Rev Lett* 108, 267201, doi:Artn 267201 Doi 10.1103/Physrevlett.108.267201 (2012).
12  Karhu, E. A. *et al.* Chiral modulations and reorientation effects in MnSi thin films. *Phys Rev B* 85, doi:Artn 094429 Doi 10.1103/Physrevb.85.094429 (2012).
13  Shibata, K. *et al.* Large anisotropic deformation of skyrmions in strained crystal. *Nat Nanotechnol* 10, 589-+, doi:10.1038/Nnano.2015.113 (2015).
14  Ritz, R. SPINTRONICS Skyrmions under strain. *Nat Nanotechnol* 10, 573-+,

doi:10.1038/nnano.2015.146 (2015).

15   Hu, Y. & Wang, B. Unified Theory of Magnetoelastic Effects in B20 compounds. *Submitted* (2016).

16   Mura, T. *Micromechanics of defects in solids*. Vol. 3 (Springer Science & Business Media, 1987).

17   Landau, L. D., Lifshitz, E. M., *Statistical physics I*.   (Pergamon Press, 1980).

18   Mura, T. Periodic distributions of dislocations. *Proceedings of the Royal Society of London. Series A. Mathematical and Physical Sciences* 280, 528-544 (1964).

19   Khachaturyan, A. Some questions concerning the theory of phase transformations in solids. *Soviet Phys. Solid State* 8, 2163-2168 (1967).

20   Iwasaki, J., Mochizuki, M. & Nagaosa, N. Current-induced skyrmion dynamics in constricted geometries. *Nat Nanotechnol* 8, 742-747, doi:DOI 10.1038/nnano.2013.176 (2013).

21   Tchoe, Y. & Han, J. H. Skyrmion generation by current. *Phys Rev B* 85, doi:Artn 174416 Doi 10.1103/Physrevb.85.174416 (2012).

22   Yu, X. Z. *et al.* Magnetic stripes and skyrmions with helicity reversals. *P Natl Acad Sci USA* 109, 8856-8860, doi:DOI 10.1073/pnas.1118496109 (2012).

23   Finazzi, M. *et al.* Laser-induced magnetic nanostructures with tunable topological properties. *Phys Rev Lett* 110, 177205, doi:Artn 177205 (2013).

24   Mochizuki, M. Spin-Wave Modes and Their Intense Excitation Effects in Skyrmion Crystals. *Phys Rev Lett* 108, doi:Artn 017601 (2012).

25   Onose, Y., Okamura, Y., Seki, S., Ishiwata, S. & Tokura, Y. Observation of magnetic excitations of Skyrmion crystal in a helimagnetic insulator Cu2OSeO3. *Phys Rev Lett* 109, 037603, doi:Artn 037603 Doi 10.1103/Physrevlett.109.037603 (2012).